\documentclass[aps,prc,preprint,tightenlines,superscriptaddress,showpacs,%
amssymb,byrevtex,nofootinbib]{revtex4}
\usepackage[dvips]{graphicx,color}
\usepackage{dcolumn}
\usepackage{epsfig}
\usepackage{bm}

\newcommand{\slas}[1]{\not\!{#1}}

\begin{document}
\title{Production of $J/\Psi$ on the nucleon and deuteron targets}
\author{ Jia-Jun Wu}
\affiliation{Physics Division, Argonne National Laboratory, Argonne, Illinois 60439, USA}
\author{T.-S. H. Lee}
\affiliation{Physics Division, Argonne National Laboratory, Argonne, Illinois 60439, USA}
\begin{abstract}
A coupled-channel model with $\pi N$, $\rho N$ and $J/\Psi N$ channels is developed to predict the  $\pi + N \rightarrow J/\Psi + N $ cross sections. The $J/\Psi$-$N$ interaction is parameterized in a form related to what has been predicted by the effective field theory approach and Lattice QCD. The other interactions within the model are constrained by the decay width of $J/\Psi \rightarrow \rho + \pi$ and the total cross section data of $\pi N$ reactions. The calculated meson-baryon amplitudes are then used to predict the cross sections of the $J/\Psi$ production on the deuteron target by including the contributions from the impulse term and  the one-loop calculations of the final $NN$ and $J/\Psi N$ re-scattering effects. Predictions of the dependence of the cross sections of $\pi^- + p \rightarrow J/\Psi+ n$, $\gamma + d \rightarrow J/\Psi + n + p$, and $\pi^+ d \rightarrow J/\Psi +p +p$ on the $J/\Psi$-$N$ potentials are presented for experimental determinations of the $J/\Psi$-$N$ interaction. Within the vector meson dominance  model, we have also applied the constructed coupled-channel model to predict the $\gamma + p\rightarrow J/\Psi + p$ cross sections near the $J/\Psi$ production threshold.
\end{abstract}

\maketitle

\section{introduction}

The $J/\Psi$ meson,
as a $c\bar{c}$ bound state,
can only interact with
the nucleon
  through the
gluon-exchange mechanism.
By using the effective field theory
method\cite{pesk79,luke92,brodsky-1,russia}
 and  Lattice QCD \cite{lqcd},
it was found that
the $J/\Psi$-$N$  interaction is attractive. Strong attraction
was also found\cite{brodsky90}
 within a Pomeron-quark coupling model of $c\bar{c}$-nucleus interactions.
Thus it is possible
that $J/\Psi$ and a system of nucleons can form nuclear bound states with
$hidden$ $charm$.
If such $exotic$ nuclear states
indeed exist and can be detected,
we can get useful information for understanding the role of gluons in nuclei.

The strength of the
$J/\Psi$-$N$ interaction, parameterized as a Yukawa form
$v_{J/\Psi N, J/\Psi N} = - \alpha \frac{e^{-\mu r}}{r}$, deduced from
Refs.\cite{pesk79,luke92,brodsky-1,russia,lqcd,brodsky90} are rather
different from each other. With $\mu= 0.60 $ GeV,
the calculated $J/\Psi$-$N$ scattering lengths
range from -0.05 fm from using $\alpha=0.06$ of
 Ref.\cite{russia} to -8.83 fm from $\alpha=0.60$ of Ref.\cite{brodsky90}.
To understand quantitatively the role of gluons in hadron interactions,
it is  important to investigate how the $J/\Psi$-$N$ interaction
can be extracted from experiments to test these theoretical models.
This is the objective of this work.
Clearly, this is also a necessary step toward searching
for possible nuclear bound states with hidden charm, as investigated
in  Ref.\cite{wulee}.

We first consider the  $\pi+ N \rightarrow J/\Psi+ N$ reaction. At  energies above
the $J/\Psi$ production threshold, this reaction is strongly influenced
 by many possible reaction
channels, such as $\rho N$, $\pi\pi N$, $\pi\pi\pi N$, $\pi\pi K\Lambda$
channels,
 because of the unitarity condition.
To proceed, we  cast this complex reaction problem
into a manageable coupled-channel model with only
$\pi N$, $\rho N$, and $J/\Psi N$ channels.
The interaction $v_{J/\Psi N, J/\Psi N}$ is chosen to be of a
form related to the Yukawa form of
Refs.\cite{pesk79,luke92,brodsky-1,russia,lqcd,brodsky90}.
With the $J/\Psi$-$\pi$-$\rho$ coupling constant determined\cite{wulee}
 by the partial decay width
of $J/\Psi \rightarrow \pi \rho$, we can calculate the $\pi N, \rho N \rightarrow J/\Psi N$
 transition potentials
$v_{\rho N, J/\Psi N}$ and $v_{\pi N, J/\Psi N}$
by using the one-$\pi$ and one-$\rho$ exchange mechanisms, respectively,
as illustrated in Fig.\ref{fg:pi-rho-exc}.
The other interactions, which do not connect with the $J/\Psi N$ channel directly, will be treated
as phenomenological complex potentials with their parameters constrained by
reproducing the
total cross sections of $\pi N$ reactions and $\pi N \rightarrow \rho N$
reactions
in the considered  energy region.
We also constrain these potentials by
reproducing
 the total cross section data of
$\gamma p \rightarrow \rho^0 p$ which can be related qualitatively to
$\rho^0 p \rightarrow \rho^0 p$
within the vector meson dominance model. To facilitate the
experimental determination of the $J/\Psi$-$N$ interaction,
we  will
 present results showing the dependence
of the calculated cross sections of $\pi^-+ p\rightarrow J/\Psi+ n$ on
the potential $v_{J/\Psi N, J/\Psi N}$.

With the meson-baryon amplitudes generated from the constructed coupled-channel model
described above and the Pomeron-exchange model of $\gamma +N \rightarrow J/\Psi+ N$ we
had developed in Ref.\cite{wulee}, we then examine the dependence
of the $\gamma + d \rightarrow J/\Psi + p + n$ cross sections on $v_{J/\Psi N, J/\Psi N}$.
We also follow the suggestion of Ref.\cite{brodsky-2} to investigate
$\pi^+ +  d \rightarrow J/\Psi + p + p$ reaction.
Instead of using the factorization
approximation used in Ref.\cite{brodsky-2}, we have performed  complete one-loop
calculations to account for both the on- and off-shell $J/\Psi N$ final state interactions.
In addition, we also include the effects due to the final $NN$ interaction by using the
scattering t-matrix generated from the Bonn potential\cite{Machleidt}.

If we use the vector meson dominance hypothesis to convert the incoming photon into
 $J/\Psi$ and $\rho$, we can predict the
$\gamma + N \rightarrow J/\Psi + N$ cross sections
within the constructed coupled-channel model. We will see that
our predictions also depend on the $J/\Psi$-$N$ potential $v_{J/\Psi N, J/\Psi N}$.
We will also present these results which could be tested
in the forthcoming experiment\cite{jpsi-jlab}.

In Section II, we present our coupled-channel model for $J/\Psi$ production
on the nucleon.The formula for calculating the meson-baryon
reaction amplitudes are given in section III.
In section IV, we give the formula for calculating  the cross sections
of the $\pi+ N \rightarrow J/\Psi+ N$ and $\gamma/\pi + d \rightarrow J/\Psi+N+N$
reactions.
Our results are presented in section V.
A summary is given in section VI.

\section{Coupled-channel model for $J/\Psi$ production}
At energies near the
 $J/\Psi$ production threshold, the $\pi N$ reactions
involve many  meson-baryon channels
such as $\pi\pi N$, $\pi\pi \pi N$, $\pi \pi K \Lambda$ etc.
Within the formulation of Ref.\cite{msl07},
the scattering $T$-matrix for such reactions
are defined
by the following coupled-channel equations
\begin{eqnarray}
T_{\alpha,\beta}(E)= v_{\alpha,\beta} + \sum_{\gamma} v_{\alpha,\gamma}
G_{\gamma}(E)T_{\gamma,\beta} \,,
\label{eq:cceq}
\end{eqnarray}
where $\alpha, \beta, \gamma$ denote the considered channels,
$G_\alpha(E)$ is the meson-baryon propagator
of channel $\alpha$, and
$v_{\alpha,\beta}$ are the interaction potentials.

For investigating the $J/\Psi$-$N$ interaction,
we will treat $\pi N$ and $\rho N$ channels explicitly since
their interactions with the $J/\Psi N$ channel can be calculated,
as will be explained later.
We thus  set
$\alpha, \beta,\gamma =J/\Psi N,  \pi N, \rho N$, $X$ in
 Eq.(\ref{eq:cceq}), where
$X$ denotes collectively all other channels.
By using the standard projection operator technique\cite{feshbach,msl07},
we can cast
 Eq.(\ref{eq:cceq}) into
\begin{eqnarray}
T_{i,j}(E)= V_{i,j}(E) + \sum_{k} V_{i,k}(E)
G_{k}(E)T_{k,j}(E)\,,
\label{eq:cceq-p}
\end{eqnarray}
where $i,j, k = J/\Psi N, \pi N, \rho N$. The energy-dependent
 interactions in Eq.(\ref{eq:cceq-p}) are
\begin{eqnarray}
V_{i,j}(E) = v_{i,j} + \sum_{X} v_{i,X}G_{X}(E)[1+t_{X,X}(E)G_X(E)]v_{X,j}\,,
\label{eq:vij}
\end{eqnarray}
where the  scattering amplitude $t_{X,X}$ in the subspace of the channel $X$
is defined by
\begin{eqnarray}
t_{X,X}(E) = v_{X,X}+ v_{X,X}G_{X}(E) t_{X,X}(E) \,.
\label{eq:txx}
\end{eqnarray}

\begin{figure}[t]
\centering
\epsfig{file=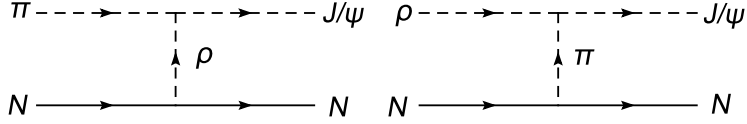, width=0.6\hsize}
\caption{Reaction mechanisms of $\pi + N \to J/\psi + N$(left) and $\rho + N \to J/\psi + N$(right).}
\label{fg:pi-rho-exc}
\end{figure}

We now turn to defining the interaction potentials
$v_{ij}$ in Eq.(\ref{eq:vij}). As noticed in Refs.\cite{wulee,brodsky-1},
 the decay width $\Gamma_{J/\Psi, \rho\pi}$ of
 $J/\Psi \rightarrow \pi \rho$ is significant.
From the value of $\Gamma_{J/\Psi, \rho\pi}$ listed by the Particle Data Group (PDG),
we  can determine\cite{wulee} the
coupling constant $g_{J/\Psi,\pi\rho}$ of the following interaction Lagrangian
\begin{eqnarray}
L_{J/\Psi \pi \rho} = -\frac{g_{J/\Psi,\pi\rho}}{m_{J/\Psi}}\epsilon^{\mu\nu\alpha\beta}
\partial_\mu \rho_\nu \partial_\alpha\phi_{J/\Psi, beta} \phi_\pi\,.
\end{eqnarray}
We find\cite{wulee} $g_{J/\Psi,\pi\rho}=0.032$.
By using  $L_{J/\Psi \pi \rho}$ and the well known\cite{sl96} Lagrangian
 $L_{\pi NN}$ and $L_{\rho NN}$,
we can calculate one-$\pi$ (one-$\rho$) exchange
transition potential of the
$\rho+ N \rightarrow J/\Psi+ N$ ($\pi + N \rightarrow J/\Psi+ N$) process,
as illustrated in Fig.\ref{fg:pi-rho-exc}.
Explicitly, their matrix elements can be written (with the normalizations
to be specified in section III and
omitting spin and isospin indices) as
\begin{eqnarray}
<\vec{p}_{J/\psi},\vec{p}_{N_f}|
v_{J/\psi N,\pi N}|\vec{p}_\pi,\;\vec{p}_{N_i}>
&=&\frac{1}{(2\pi)^3}\frac{1}{\sqrt{2E_{J/\Psi}(\vec{p}_{J/\psi})}}
\sqrt{\frac{m_N}{E_N(\vec{p}_{N_f})}}\sqrt{\frac{m_N}{E_N(\vec{p}_{N_i})}}
\frac{1}{\sqrt{2E_{\pi}(\vec{p}_\pi)}}\nonumber \\
&&\times\left(-\frac{g_{J/\psi\rho\pi}}{m_{J/\psi}}g_{\rho NN}\right)
\epsilon_{\mu\nu\alpha\beta}q^\mu p_{J/\psi}^\alpha \varepsilon^{\beta}(p_{J/\Psi})
\frac{1}{q^2-m_\rho^2}\nonumber\\
&&\times\bar{u}_{\vec{p}_{N_f}}
\left[\gamma^\nu-\frac{q\!\!\!/ q^\nu}{m^2_\rho}
+\frac{\kappa_\rho}{4m_N}(\gamma^{\nu}q\!\!\!/-q\!\!\!/\gamma^{\nu})\right]
u_{\vec{p}_{N_i}},
\label{eq:pinjn} \\
&& \nonumber \\
<\vec{p}_{J/\psi},\vec{p}_{N_f}| v_{J/\psi N,\rho N}
|\vec{p}_\rho,\vec{p}_{N_i}>
&=&\frac{-1}{(2\pi)^3}\frac{1}{\sqrt{2E_{J/\Psi}(\vec{p}_{J/\psi})}}
\sqrt{\frac{m_N}{E_N(\vec{p}_{N_f})}}\sqrt{\frac{m_N}{E_N(\vec{p}_{N_i})}}
\frac{1}{\sqrt{2E_{\rho}(\vec{p}_\rho)}}\nonumber \\
&\times&\left(\frac{g_{J/\psi\rho\pi}}{m_{J/\psi}}\frac{f_{\pi NN}}{m_\pi}\right)
\epsilon_{\mu\nu\alpha\beta}p_{\rho}^\mu \varepsilon^\nu(p_\rho) p_{J/\psi}^\alpha
\varepsilon^{\beta}(p_{J\Psi}) \nonumber \\
&&\times \frac{1}{q^2-m_\pi^2}
\bar{u}_{\vec{p}_{N_f}}
q\!\!\!/ \gamma^5
u_{\vec{p}_{N_i}} \,,
\label{eq:rhonjn}
\end{eqnarray}
where ${q}={p}_{N_i}-{p}_{N_f}$, the coupling constants
$g_{\rho NN}=6.20$, $\kappa_\rho=1.825$, $f_{\pi NN}=\sqrt{4\pi \times 0.80}$
are taken from a dynamical model\cite{sl96} of $\pi N$ scattering.
All external particles of the matrix element of the above equations
 are on their mass shell with
their  momenta defined as $p_a=(E_a(\vec{p}_a), \vec{p}_a)$
and $E_a(\vec{p}_a) = [m^2_a+\vec{p}^{\,\, 2}_a]^{1/2}$.

We assume that the  mesons and baryons in the channel $X$ do not contain
charmed quarks. The coupling
between $X$ and $J/\Psi N$ can then be neglected and we can set
$v_{J/\Psi N, X} =0$. It follows that for the interactions involving
$J/\Psi N$,
the second term of Eq.(\ref{eq:vij}) vanishes and we have the following
simplification,
\begin{eqnarray}
V_{\pi N, J/\Psi N} &=& v_{\pi N, J/\Psi N}\,, \label{eq:vi0} \\
V_{\rho N, J/\Psi N} &=& v_{\rho N, J/\Psi N}\,, \\
V_{J/\Psi N, J/\Psi N} &=& v_{J/\Psi N, J/\Psi N}\,.
\label{eq:vi0}
\end{eqnarray}
The matrix elements for $v_{\pi N, J/\Psi N}$ and $v_{\rho N, J/\Psi N}$ have been
given in Eqs.(\ref{eq:pinjn}) and (\ref{eq:rhonjn}).
For $v_{J/\Psi N, J/\Psi N}$, we follow
 Refs.\cite{luke92,brodsky-1,russia,lqcd} and use the following Yukawa form
\begin{eqnarray}
v_{J/\Psi N, J/\Psi N}(r) =  -\alpha \frac{e^{-\mu_0 r}}{ r} \,.
\label{eq:vjpnjpn}
\end{eqnarray}
To be consistent with the relativistic expression
 Eqs.(\ref{eq:pinjn}) and (\ref{eq:rhonjn}), we assume
that Eq.(\ref{eq:vjpnjpn}) is the non-relativistic limit of
a one-scalar meson exchange
amplitude in field theory. We then obtain
\begin{eqnarray}
<\vec{p}_{J/\psi}^{\,\,'},\vec{p}_{N_f}| v_{J/\Psi N ,J/\psi N}
|\vec{p}_{J/\Psi},\vec{p}_{N_i}>
&=&\frac{1}{(2\pi)^3}\frac{1}{\sqrt{2E_{J/\Psi}(\vec{p}_{J/\psi}^{\,\,'})}}
\sqrt{\frac{m_N}{E_N(\vec{p}_{N_f})}}\sqrt{\frac{m_N}{E_N(\vec{p}_{N_i})}}
\frac{1}{\sqrt{2E_{J\Psi}(\vec{p}_{J/\Psi})}}\nonumber \\
&\times&[g_{\nu\mu}\varepsilon^\nu({p}_{J/\psi}')
\varepsilon^\mu({p}'_{J/\psi})]
\frac{V_0}{q^2-\mu^2_0}
\bar{u}_{\vec{p}_{N_f}}
u_{\vec{p}_{N_i}}\,,
\label{eq:jpnjpn}
\end{eqnarray}
where $V_0= -8\alpha \pi m_{J/\Psi}$, and $q={p}_{N_i}-{p}_{N_f}$.

Since $v_{\pi N, X}$, $v_{\rho N,X}$ and $t_{XX}(E)$ can not be calculated theoretically, we
will determine $V_{i,j}(E)$ with $i,j= \pi N, \rho N$  in Eq.(\ref{eq:vij}) phenomenologically.
For simplicity, we  assume that they all have the following local form in
the non-relativistic limit
\begin{eqnarray}
V_{i,j}(E) = v^0_{i,j}(E) f_{i,j}(\vec{r}) \,\,\,\,;\,\,\, i,j=\pi N, \rho N\,,
\label{eq:v-fr}
\end{eqnarray}
where $f_{i,j}(\vec{r})$ is unitless such as $f_{i,j}(\vec{r}) = 1/(1+e^{(r-r_0)/t})$
or $e^{-r^2/b^2}$. In the strong absorption (diffractive) model\cite{feshbach}, the form of
$f_{i,i}(\vec{r})$ for $i=1,2 $
is similar to
the nucleon density. We will specify  $f_{i,j}(\vec{r})$ later.

We next define
\begin{eqnarray}
F_{i,j}(\vec{q},E)= v^0_{i,j}(E) \int
e^{-i\vec{q}\cdot \vec{r}} f_{i,j}(\vec{r})d\vec{r}\,.
\end{eqnarray}
Including appropriate covariant spin factors, the matrix elements corresponding
to the form of Eq.(\ref{eq:v-fr}) are taken to be
\begin{eqnarray}
<\vec{p}_\pi',\vec{p}_{N_f}| V_{\pi N ,\pi N}(E)
|\vec{p}_\pi,\vec{p}_{N_i}>
&=&\frac{1}{(2\pi)^3}
\bar{u}_{\vec{p}_{N_f}}\; F_{\pi N,\pi N}(\vec{q},E)\; u_{\vec{p}_{N_i}} \,,
\label{eq:pnpn} \\
<\vec{p}_{\rho}',\vec{p}_{N_f}| V_{\rho N ,\rho N}(E)
|\vec{p}_{\rho},\vec{p}_{N_i}>
&=&\frac{1}{(2\pi)^3}
\varepsilon^\nu_{\rho} ({p}_{\rho}')\bar{u}_{\vec{p}_{N_f}}\;
F_{\rho N, \rho N}(\vec{q},E) \;
u_{\vec{p}_{N_i}}\varepsilon_{\rho,\nu}({p}_{\rho})\,,
\label{eq:rnrn}  \\
<\vec{p}_{\rho},\vec{p}_{N_f}|
V_{\rho N,\pi N}(E)|\vec{p}_\pi,\;\vec{p}_{N_i}>
&=&\frac{1}{(2\pi)^3}
\frac{1}{m_\pi}p^\mu_\pi  \varepsilon_\mu(p_\rho)\;
F_{\rho N,\pi N}(\vec{q},E)\;\bar{u}_{\vec{p}_{N_f}}
[\frac{1}{m_\pi}\gamma_5\slas{q}]u_{\vec{p}_{N_i}}.
\label{eq:rnpn}
\end{eqnarray}
We will determine the parameters $v^0_{i,j}$ and $f_{i,j}(\vec{r})$
 phenomenologically
by fitting the total cross sections of $\pi N$ and $\pi N \rightarrow \rho N$, and
the $\gamma N\rightarrow \rho N $ cross sections which can be calculated from
the predicted $\rho N\rightarrow \rho N $ amplitudes using the vector meson dominance hypothesis.

\section{Calculations of Reaction Amplitudes}

In this work we follow the formulation
of  Ref.\cite{msl07} within which the scattering $T$-matrix is related to
the $S$-matrix
by
\begin{eqnarray}
S_{f,i}(E) = \delta_{f,i} - (2\pi) i\delta(E_f-E_i)\delta(\vec{P}_f-\vec{P}_i) T_{f,i}(E)\,,
\label{eq:s-t}
\end{eqnarray}
where $E_\alpha$ and $\vec{P}_\alpha$ are the total energy and momentum of the state $\alpha$.
The normalizations for the plane wave state $|\vec{k}>$ and
bound state $|\Phi_a>$  are defined by
\begin{eqnarray}
<\vec{k}|\vec{k}^{\,\,'}> &=& \delta(\vec{k}-\vec{k}^{\,\,'})\,,
\label{eq:nor-plane} \\
<\Phi_a|\Phi_b> &=& \delta_{ab}\,.
\label{eq:nor-bound}
\end{eqnarray}

\subsection{Meson-Baryon reaction amplitudes}
In our calculations, we need
the meson-baryon ($MB$) matrix elements in a fast moving frame.
Following the instant form of
relativistic quantum mechanics\cite{polyzou}, we write for the two-particle
$M(\vec{p}_M)+B (\vec{p}_B) \rightarrow M'(\vec{p}^{\,\,'}_M)+
B'(p^{\,\,'}_B)$
transition
\begin{eqnarray}
&&<\vec{p}^{\,\,'}_M m'_{j_M},  m'_{i_M};
\vec{p}^{\,\,'}_B m'_{j_B}, m_{\tau_B}|{T}_{M'B',MB}(E)|\vec{p}_M m_{j_M}, m_{i_M};
\vec{p}_B, m_{j_B} m_{\tau_B}> \nonumber \\
&&=
J(\vec{P},\vec{k};\vec{p}_M,\vec{p}_B)
J(\vec{P}^{\,\,'},\vec{k}^{\,\,'};\vec{p}^{\,\,'}_M,\vec{p}^{\,\,'}_B) \nonumber \\
&&\times
< m'_{j_M},  m'_{i_M};
 m'_{j_B}, m'_{\tau_B}|t_{M'B',MB}(\vec{k}^{\,\,'},\vec{k};W)|m_{j_M},m_{i_M};
 m_{j_B}, m_{\tau_B}> \,,
\label{eq:cm-lab}
\end{eqnarray}
where $[ m_{j_M},m_{i_M}]$
and $[m_{j_B},m_{\tau_B}] $ are the z-components of the  spin-isospin quantum numbers
for  $M$ and $B$, respectively.
 The total momenta $\vec{P}, \vec{P}^{\,\,'}$ and
the energy $W$  in the center of mass system are
\begin{eqnarray}
\vec{P}&=&\vec{p}_M+\vec{p}_B\,, \nonumber \\
\vec{P}^{\,\,'}&=& \vec{p}^{\,\,'}_M+\vec{p}^{\,\,'}_B\,, \nonumber \\
W &=& [E^2- \vec{P}^{\,\,2}]^{1/2}\,.
\label{eq:cm-e}
\end{eqnarray}
The relative momentum $\vec{k}$ and Jacobian $J(\vec{P},\vec{k};\vec{p}_M,\vec{p}_B)$
in Eq.(\ref{eq:cm-lab}) are defined by the Lorentz boost
transformation\cite{polyzou}
\begin{eqnarray}
&&\vec{k} = \vec{p}_M+\frac{\vec{P}}{M_0}[\frac{\vec{P}\cdot \vec{p}_M}{M_0+H_0}
-E_{M}(\vec{p}_M)]\,,
\label{eq:rel-k} \\
&&J(\vec{P},\vec{k};\vec{p}_M,\vec{p}_B) =|\frac{\partial(\vec{P},\vec{k})}
{\partial(\vec{p}_M,\vec{p}_B)}|^{1/2}
=[\frac{E_{M}(\vec{k})E_{B}(\vec{k})H_0}
{E_M(\vec{p}_M)E_B(\vec{p}_B)M_0}]^{1/2}\,,
\label{eq:jcob}
\end{eqnarray}
with
\begin{eqnarray}
H_0&=&E_{M}(\vec{p}_M)+E_{B}(\vec{p}_B)\,, \\
M_0 &=&\sqrt{H^2_0-\vec{P}^{\,\,2}} \nonumber \\
&=&E_{M}(\vec{k})+E_{B}(\vec{k})\,.
\label{eq:h0m0})
\end{eqnarray}

In Eq.(\ref{eq:cm-lab}), $t_{M'B',MB}(W)$ is the scattering operator in the
center of mass frame.
In the partial-wave
representation, we can write
\begin{eqnarray}
t_{M'B',MB}(\vec{k}^{\,\,'},\vec{k};W)
&=&\sum_{JM_J,TM_T}\sum_{L'S',LS}
|{y}^{JM_J,TM_T}_{L'S',M'B'}(\hat{k}^{\,\,'})>
t^{JT}_{L'S',M'B',LS,MB}(k',k,W)
<{ y}^{JM_J,TM_T}_{LS,MB}(\hat{k})| \nonumber \\
&&\label{eq:pwa}
\end{eqnarray}
where the angle-spin-isospin vector is defined by
\begin{eqnarray}
|{y}^{JM_J,TM_T}_{LS,MB}(\hat{k})>
&=&\sum_{m_{j_B}m_{j_M}}\sum_{m_{\tau_B}m_{i_M}}
| m_{j_M},  m_{i_M};
 m_{j_B},  m_{\tau_B}> \nonumber \\
&&\times [\sum_{m_L,m_S} <J\,\,M_J|L\,\,S\,\,m_L\,\,m_S>
<S\,\, m_s|j_M\,\,j_B\,\,m_{j_M},m_{j_B}> \nonumber \\
&&\times
<T\,\,M_T|i_M\,\,\tau_B\,\, m_{i_M}\,\,m_{\tau_B}>
 Y_{L\,\,m_L}(\hat{k})]\,. \nonumber \\
&& \label{eq:s-i-a}
\end{eqnarray}
Here $[j_M, i_M]$ and $[j_B, \tau_B]$ are the spin-isospin quantum numbers for
$M$ and $B$, respectively; $<jm|j_1j_2m_1m_2>$ is the Clebsch-Gordon coefficient and
$Y_{L\,\,m_L}(\hat{k})$ the spherical harmonic function, as defined in
Ref.\cite{brink}.
With the same partial-wave expansion Eq.(\ref{eq:s-i-a}) for the meson-baryon potential $V_{i,j}(E)$,
the  Eqs.(\ref{eq:cceq-p}) and (\ref{eq:cm-lab}) lead to
the following coupled-channel equation
\begin{eqnarray}
&&t^{JT}_{L'S',M'B',LS,MB}(k',k.W) =
V^{JT}_{L'S',M'B',LS,MB}(k',k.E) \nonumber \\
&&+ \sum_{L^{''}S^{''},M^{''}B^{''} }
\int k^{'' 2} dk^{''}
V^{JT}_{L'S',M'B',L^{''}S^{''},M^{''}B^{''}}(k',k^{''}.E)
G_{M^{''}B^{''} }(k^{''},W)t^{JT}_{L^{''}S^{''},M^{''}B^{''},LS,MB}(k',k.W)
\nonumber \\
&& \label{eq:cceq-pw}
\end{eqnarray}
where $MB, M'B', M^{''}B^{''} = \pi N, \rho N, J/\Psi N$, and the propagator is
\begin{eqnarray}
G_{MB}(k,W)=\frac{1}{W-E_M(k)-E_B(k)+i\epsilon }\,. \nonumber
\end{eqnarray}
We use the  procedures developed in Ref.\cite{msl07} to
 calculate  the potential matrix elements $V^{JT}_{L'S',M'B',L^{''}S^{''},M^{''}B^{''}}(k',k^{''}.E)$ from the matrix elements
Eqs.(\ref{eq:pinjn})-(\ref{eq:rnpn}).

\subsection{Deuteron wavefunction and $NN$ amplitude}
For the calculations on the deuteron target,
we need
the deuteron bound state $|\Psi_{\vec{p}_d,M_{J_d}}>$ moving
 with a high
momentum  $\vec{p}_d$. Following Ref.\cite{polyzou}, it is
 defined by
\begin{eqnarray}
&&<\vec{p}_1,m_{s_1}m_{\tau_1};\vec{p}_2,m_{s_2}m_{\tau_2}|\Psi_{\vec{p}_d,M_d}>\nonumber \\
 && =
\delta(\vec{p}_d-\vec{p}_1-\vec{p}_2)
<\vec{p}_1,m_{s_1}m_{\tau_1};\vec{p}_2,m_{s_2}m_{\tau_2}|\Phi_{\vec{p}_d,M_d}> \,,
\end{eqnarray}
where
\begin{eqnarray}
<\vec{p}_1,m_{s_1}m_{\tau_1};\vec{p}_2,m_{s_2}m_{\tau_2}|\Phi_{\vec{p}_d,M_d}>
=  J(\vec{p}_d,\vec{\kappa};\vec{p}_1,\vec{p}_2)
 <m_{s_1}m_{\tau_1};m_{s_2}m_{\tau_2}|\chi^{J_d M_d,T_d M_{T_d}}(\vec{\kappa})>
\label{eq:deut-wf}
\end{eqnarray}
with
\begin{eqnarray}
|\chi^{J_d M_d,T_dM_{T_d}}(\vec{\kappa})> = \sum_{l=0,2}
|y^{J_dM_d,T_d M_{T_d}}_{ls_d,NN} (\hat{\kappa})> u_l(\kappa) \,.
\end{eqnarray}
Here $s_d=1$ is the deuteron spin and $u_l(\kappa)$ is the usual deuteron radial
wave function in its rest frame.
The Jacobian $J(\vec{p}_d,\vec{\kappa};\vec{p}_1,\vec{p}_2)$
 and the relative momentum
$\vec{\kappa}$ can be calculated by using the same
 Eqs.(\ref{eq:rel-k})-(\ref{eq:h0m0})
with the replacement $\vec{k}\rightarrow \vec{\kappa}$,
$\vec{p}_M \rightarrow \vec{p}_1$, $\vec{p}_B \rightarrow \vec{p}_2$,
$\vec{P} \rightarrow \vec{p}_d$,  and
$MB\rightarrow NN$.

We also need the $NN$ amplitude
\begin{eqnarray}
&& <\vec{p}^{\,\,'}_1 m'_{j_1},  m'_{\tau_1};
\vec{p}^{\,\,'}_2 m'_{j_2}, m_{\tau_2}|{T}_{NN,NN}(E)|\vec{p}_1 m_{j_1}, m_{i_1};
\vec{p}_2, m_{j_2} m_{\tau_2}> \nonumber \\
&&=
J(\vec{P},\vec{k};\vec{p}_1,\vec{p}_2)
J(\vec{P}^{\,\,'},\vec{k}^{\,\,'};\vec{p}^{\,\,'}_1,\vec{p}^{\,\,'}_2) \nonumber \\
&&\times
< m'_{j_1},  m'_{\tau_1};
 m'_{j_2}, m'_{\tau_2}|t_{NN,NN}(\vec{k}^{\,\,'},\vec{k};W)|m_{j_1},m_{\tau_1};
 m_{j_2}, m_{\tau_2}> \,.
\label{eq:cm-lab-NN}
\end{eqnarray}
The above matrix element can be calculated from the same equations
Eqs.(\ref{eq:cm-lab})-(\ref{eq:s-i-a}) with the replacement of
$MB \rightarrow NN$. We generated the $NN$ partial-wave matrix elements
$t^{JT}_{L'S'NN,LSNN}(k',k;W)$ from the Bonn potential\cite{Machleidt}.

\section{Calculations of cross sections}

\subsection{Cross sections of  Meson-baryon reactions}

We need to evaluate the cross sections for the reactions
involving three
meson-baryon ($MB$) channels with
$MB = \pi N, \rho N, J/\Psi N$.
With the definitions Eqs.(\ref{eq:s-t})-(\ref{eq:cm-lab}),
the differential cross sections in the center of mass (CM)
for the $M(\vec{k}) + B(-\vec{k}) \rightarrow M'(\vec{k}^{\,\,'})
+ B'(-\vec{k}^{\,\,'})$ reaction can be written as
\begin{eqnarray}
\frac{d\sigma}{d\Omega} &=& \frac{(4\pi)^2}{k^2}
\frac{\rho_{M'B'}(k')\rho_{MB}(k)}{(2j_M+1)(2j_B+1)}
\sum_{m_{j_M},m_{j_B}}\sum_{m^\prime_{j_M},m^\prime_{j_B}}[ \nonumber \\
&&| < m'_{j_M}, m'_{i_M};
 m'_{j_B},  m'_{\tau_B}|t_{M'B',MB}(\vec{k}^{\,\,'},\vec{k};W)| m_{j_M},  m_{i_M};
 m_{j_B},  m_{\tau_B}>|^2\,\, ]\;,\nonumber \\
&&
\label{eq:dcrst-mb}
\end{eqnarray}
where the matrix element of $t_{M'B',MB}(\vec{k}^{\,\,'};W)$ can be
calculated from Eqs.(\ref{eq:pwa})-(\ref{eq:s-i-a}) and the solution of
Eq.(\ref{eq:cceq-pw}). The density of state  is defined by
\begin{eqnarray}
\rho_{MB}(k) = \pi \frac{kE_M(k)E_B(k)}{E}\;.
\end{eqnarray}
The total cross sections $\sigma^{el}_{\pi N, \pi N}$ of elastic $\pi N\rightarrow \pi N$ and
$\sigma_{\pi N, \rho N}$ of $\pi N \rightarrow \rho N$ can be calculated from
using Eq.(\ref{eq:dcrst-mb}), Eq.(\ref{eq:pwa}) and
the solution of the coupled-channel equation Eq.(\ref{eq:cceq-pw}).

\subsection{Total cross sections of $\pi N$ reactions}
The total cross sections $\sigma^{tot}_{\pi N}$
 can be obtained from the $\pi N$ elastic scattering amplitude using
the optical theorem. With our definitions
Eqs.(\ref{eq:s-t})-(\ref{eq:cm-lab}), we have
\begin{eqnarray}
\sigma^{tot}_{\pi N} = -Im\, [\bar{f}_{\pi N,\pi N}(\theta=0)]
\end{eqnarray}
where the spin-averaged $\pi N$ elastic scattering amplitude is
\begin{eqnarray}
\bar{f}_{\pi N,\pi N}(\theta) &=&\frac{(4\pi)^2}{k^2}\frac{\rho_{\pi N}(E) }{2j_N+1}
 [\sum_{m_{j_N}}
<0, m_{i_\pi};
 m_{j_N}, m_{\tau_N}|t_{\pi N,\pi N}(\vec{k}^{\,\,'},\vec{k};E)
|0, m_{i_\pi};
 m_{j_N}, m_{\tau_N}>] \nonumber \\
&&
\end{eqnarray}
and
 $k=|\vec{k}|=|\vec{k}^\prime|$
is the on-shell
momentum, and $cos\theta = \hat{k}^{\,\,'}\cdot \hat{k}$.

\subsection{Cross sections of photo-production of $J/\Psi$ and $\rho$}
We now note that
with the vector meson dominance hypothesis, we can predict the cross section of
$\gamma + p \rightarrow J/\Psi+ p$
within the constructed coupled-channel model.
This is done by
writing the amplitude for
$\gamma(\vec{q})+ p(-\vec{q}) \rightarrow J/\Psi(\vec{k}^{\,\,'})
+p(-\vec{k}^{\,\,'})$ as
\begin{eqnarray}
&&< m'_{J/\Psi}, m'_{j_N}|
t_{J/\Psi p, \gamma p}(\vec{k}^{\,\,'},\vec{q};W)|
\lambda_{\gamma}, m_{j_N}> \nonumber \\
&&
=\sum_{m_{J/\Psi}}< m'_{J/\Psi}, m'_{j_N}|
t_{J/\Psi p, J/\Psi p}(\vec{k}^{\,\,'},\vec{k};W)|
m_{J/\Psi}, m_{j_N}>\frac{e}{f_{J/\Psi}}\delta_{m_{J/\Psi},\lambda_{\gamma}}
\nonumber \\
&&
+\sum_{m_{\rho}}< m'_{J/\Psi}, m'_{j_N}|
t_{J/\Psi p, \rho^0 p}(\vec{k}^{\,\,'},\vec{k};W)|
m_{\rho^0}, m_{j_N}>\frac{e}{f_{\rho}}\delta_{m_{\rho^0},\lambda_{\gamma}}\,,
\label{eq:gn-jn}
\end{eqnarray}
where $f_{J/\Psi}=11.2$ and $f_\rho=5.33$.

Obviously we can get the
$\gamma(\vec{q})+ p(-\vec{q}) \rightarrow \rho^0(\vec{k}^{\,\,'})
+p(-\vec{k}^{\,\,'})$ amplitude from
Eq.(\ref{eq:gn-jn}) by interchanging $J/\Psi$ and $ \rho^0$.
This will allow us to calculate  the total cross section
 $\sigma_{\gamma p,\rho^0 p}$ of
 $\gamma p \rightarrow \rho^0 p$. Clearly, within this model based on the
vector meson dominance hypothesis,
$\sigma_{\gamma p,\rho^0 p}$ is closely related to the total cross section
of $\rho^0 p \rightarrow \rho^0 p$ which can not be obtained experimentally, but
can  be essential in determining the parameters associated with our phenomenological
potential $V_{\rho N, \rho N}(E)$.

\begin{figure}[ht]
\centering
\epsfig{file=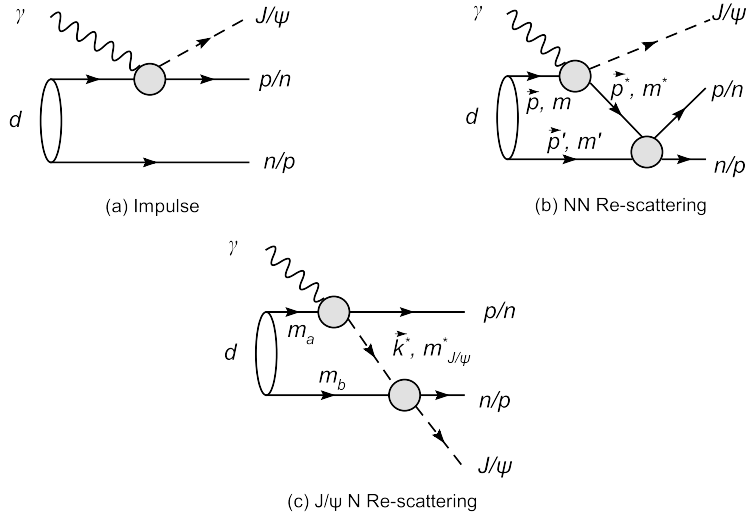, width=0.6\hsize}
\caption{Graphical representation of $J/\psi$ photo-production
on the deuteron. Graphs(a), (b), and (c) correspond, respectively,
to the impulse contribution, Eq.(\ref{eq:impulse});
$N N$ re-scattering, Eq.(\ref{eq:NN});
and $J/\psi N$ re-scattering, Eq.(\ref{eq:jpsiN}).}
\label{fg:diag}
\end{figure}

\subsection{Cross sections for $\gamma + d \rightarrow J/\Psi + n+p$ }
In Fig.\ref{fg:diag}, we illustrate the mechanisms included in
our calculations of the cross sections of
$J/\Psi$ production on the deuteron target.
Since we are mainly
interested in the $J/\Psi$-$N$ interaction (Fig.\ref{fg:diag}.(c)),
we will
examine how  the predicted cross sections
depend on the
$J/\Psi$-$N$ relative momentum in the final $J/\Psi + n+p$ state.
This can be done most effectively by
 considering  the following
 differential cross section in the CM frame of the
$\gamma(\vec{q}) + d(-\vec{q}) \rightarrow J/\Psi(\vec{k})+ n(\vec{p}_n)
+ p(\vec{p}_p)$ reaction
\begin{eqnarray}
\frac{d^2\sigma}{d\Omega_p d|\vec{\kappa}_{J/\psi}|}&=&(2\pi)^4\frac{E_\gamma(\vec{q})E_d(\vec{q})}{|\vec{q}|W}
\int d\Omega_{\hat{{\vec{\kappa}}}_{J/\psi}} \vec{\kappa}^2_{J/\psi}
\frac{E_n(\vec{p}_n)E_{J/\psi}(\vec{k})E_{p}(\vec{p}_p)}{E_n(-\vec{\kappa}_{J/\psi})E_{J/\psi}(\vec{\kappa}_{J/\psi})}
\frac{M_0|\vec{p}_p|}{W}\nonumber\\
&&\frac{1}{2(2J_{ d}+1)}\sum_{\lambda_{\gamma},\;m_{ d}}\
\sum_{m_{J/\psi},\;m_n,\;m_p}\nonumber\\
&&|<\vec{k}\,m_{J/\psi},\;\vec{p}_p\;m_p,\;\vec{p}_n\;m_n|
\hat{T}^{Imp}+\hat{T}^{J/\psi N}+\hat{T}^{N N}
|\vec{q}\;\lambda_{\gamma},\;\Phi_{-\vec{q}\,m_{d}}>|^2,
\label{eq:diffcrsec}
\end{eqnarray}
where $\vec{q}$ is the photon momentum chosen to be in the quantization $z$-direction,
$\Phi_{\vec{p}_d,m_d}$ is the deuteron with momentum
$\vec{p}_d = -\vec{q}$,
$\vec{p}_n$, $\vec{p}_p$
and $\vec{k}$ are the
 momenta of the
final   proton, neutron and $J/\psi$, respectively.
 $\vec{\kappa}_{J/\psi}=(|\vec{\kappa}_{J/\psi}|, \Omega_{\kappa_{J/\psi}})$ is
the  momentum of $J/\Psi$ in the CM of the
 $J/\Psi$-$N$ subsystem.
The amplitudes $\hat{T}^{Imp}$, $\hat{T}^{NN}$, and $\hat{T}^{J/\Psi N}$
 are calculated from the impulse ((a)), $NN$ re-scattering ((b)), and
$J/\Psi$-$N$ re-scattering ((c))
mechanisms shown in Fig.\ref{fg:diag}.

In Eq.(\ref{eq:diffcrsec}), $W$ is the invariant mass of $\gamma d$ and
$M_0 = E_{J/\Psi}(\vec{\kappa}_{J/\psi})+E_{n}(\vec{\kappa}_{J/\psi})$
is the
invariant mass of  the $J/\psi$-$N$ system.
 The magnitude of the outgoing proton momentum
$\vec{p}_p$ can be calculated from$M_0$
and $W$ by
\begin{eqnarray}
|\vec{p}_p| = \frac{1}{2W}[(W^2-m_p^2-M^2_0)^2 - 4m_p^2M^2_0]^{1/2}\,.
\label{eq:pp}
\end{eqnarray}
The direction of $\vec{p}_p$ is specified by $\Omega_p$ with respect to the incident
photon momentum $\vec{q}$.
The other variables $\vec{k}$ for the outgoing $J/\Psi$
and $\vec{p}_n$ for the outgoing neutron in Eq.(\ref{eq:diffcrsec})
can then be calculated
from $\vec{\kappa}_{J/\psi}$ and $\vec{p}_p$ as follows:
\begin{eqnarray}
\vec{k}&=&\vec{\kappa}_{J/\psi}+\frac{\vec{p}_p}{M_0}
\left(\frac{\vec{p}_{p}\cdot\vec{\kappa}_{J/\psi}}{M_0+W-E_{p}(\vec{p}_p)}-E_{J/\psi}(\vec{\kappa}_{J/\psi})\right),\\
\vec{p}_n&=&-\vec{\kappa}_{J/\psi}+\frac{\vec{p}_p}{M_0}
\left(-\frac{\vec{p}_p\cdot\vec{\kappa}_{J/\psi}}{M_0+W-E_{p}(\vec{p}_p)}-E_{n}(\vec{\kappa}_{J/\psi})\right).
\label{eq:momtna}
\end{eqnarray}
With Eqs.(\ref{eq:pp})-(\ref{eq:momtna}), all kinematic variables for calculating
the
integrand of
Eq.(\ref{eq:diffcrsec}) are completely fixed for a given $\Omega_p$ and $\vec{\kappa}_{J/\psi}$.
Our task is to explore at what $\Omega_p$, the calculated differential
cross section $\frac{d^2\sigma}{d\Omega_p d|\vec{\kappa}_{J/\psi}|}$
is most sensitive to the amplitude $\hat{T}^{J/\psi N}$ for the
$J/\Psi$-$N$ re-scattering mechanism (c) of Fig.\ref{fg:diag}.

In the following three subsections, we give formula for evaluating the matrix elements
of $T^{Imp}$, $T^{NN}$, and $T^{J/\Psi N}$.
We have evaluated the Clebsch-Gordon coefficients associated with the isospin
quantum numbers in Eq.(\ref{eq:cm-lab}) and thus all isospin indices are suppressed and
the amplitudes are on specific charged states specified explicitly as $n$ for neutron and
$p$ for proton etc.

\subsubsection{Impulse Amplitude}
\begin{figure}[h]
\centering
\epsfig{file=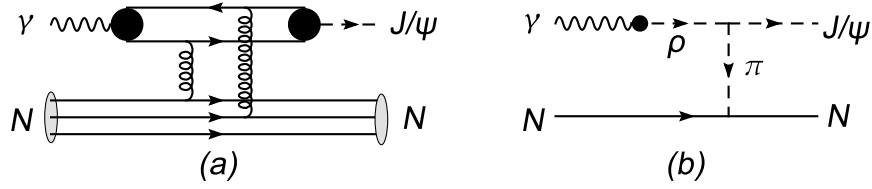, width=0.7\hsize}
\caption{Reaction mechanisms of $\gamma + N \to J/\psi + N$: (a) Pomeron-exchange, (b) pion-exchange.}
\label{fg:photon}
\end{figure}

The impulse amplitude of Eq.(\ref{eq:diffcrsec})(Fig.\ref{fg:diag}(a))
can be straightforwardly written
as
\begin{eqnarray}
&&<\vec{k}\;m_{J/\psi},\;\vec{p}_p\;m_p,\;\vec{p}_n\;m_n|
\hat{T}^{Imp}|\vec{q}\;\lambda_{\gamma},\;\Phi_{-\vec{q},m_{d}}>\nonumber\\
&=&\sum_{m_a}\left[<\vec{k}\;m_{J/\psi},\;\vec{p}_p\;m_p|
T_{J/\psi p,\gamma p}|\vec{q}\;\lambda_{\gamma},\;-\vec{q}-\vec{p}_n\;m_{a}>
<-\vec{q}-\vec{p}_n\;m_{a},\;\vec{p}_n\;m_n|
{\Phi}_{-\vec{q}\;m_{d}}>\right.\nonumber\\
&&+\left.<\vec{k}\;m_{J/\psi},\;\vec{p}_n\;m_n|
T_{J/\Psi n,\gamma n}|\vec{q}\;\lambda_{\gamma},\;-\vec{q}-\vec{p}_p\;m_{a}>
<\vec{p}_p\;m_p,\;-\vec{q}-\vec{p}_p\;m_{a}
|\Phi_{-\vec{q}\;m_{d}}>\right]\,,\nonumber\\
\label{eq:impulse}
\end{eqnarray}
where $<\vec{p}_p;m_p,\;\vec{p}_n;m_n|\Phi_{\vec{p}_d;m_{d}}>$ has been
 defined by Eq.(\ref{eq:deut-wf}) (omitting isospin indices).
The $\gamma + N_i \to J/\psi + N_f$ amplitudes in the
above expression are taken from our previous
 work\cite{wulee}
\begin{eqnarray}
&&<\vec{k}\;m_{J/\Psi},\;\vec{p}_{N_f}\;m_{N_f}|
T_{J/\Psi N,\gamma N}
|\vec{q}\;\lambda_\gamma,\;\vec{p}_{N_i}\;m_{N_i}>\nonumber\\
&=&\frac{1}{(2\pi)^3}\frac{1}{\sqrt{2E_{J/\Psi}(\vec{k})}}
\sqrt{\frac{m_N}{E_N(\vec{p}_{N_f})}}\sqrt{\frac{m_N}{E_N(\vec{p}_{N_i})}}\frac{1}{\sqrt{2|\vec{q}|}}\nonumber \\
&&\times
[\bar{u}_{m_{N_f}}(\vec{p}_{N_f})
\epsilon^*_\mu(k,m_{J/\Psi})
(\mathcal{M}^{\mu\nu}_\mathbb{P}(\vec{k},\vec{p}_{N_f},\vec{q},\vec{p}_{N_i})
+\mathcal{M}^{\mu\nu}_\pi(\vec{k},\vec{p}_{N_f},\vec{q},\vec{p}_{N_i})]
\epsilon_\nu(\vec{q},\lambda_\gamma)
u_{m_{N_i}}(\vec{p}_{N_i}))\,, \label{eq:tot-amp} \nonumber \\
&&
\label{eq:photonjpsin},
\end{eqnarray}
where $\mathcal{M}^{\mu\nu}_\mathbb{P}$
and $\mathcal{M}^{\mu\nu}_\pi$ are the contributions
from the Pomeron
exchange and $\pi$ exchange mechanisms, respectively,  as shown in the Fig.\ref{fg:photon}.

For the Pomeron exchange, the amplitude can be written as:
\begin{eqnarray}
\mathcal{M}^{\mu\nu}_\mathbb{P}(k,p_{N_f},q,p_{N_i})&=& G_\mathbb{P}(s,t)
\mathcal{T}^{\mu\nu}_\mathbb{P}(t,q),
\label{eq:MP}
\end{eqnarray}
with
\begin{eqnarray}
G_\mathbb{P}(s,t) &=& \left(\frac{s}{s_0}\right)^{\alpha_P(t)-1}
\exp\left\{ - \frac{i\pi}{2} \left[ \alpha_P(t)-1 \right]
\right\} \,,\label{eq:regge-g}\\
\mathcal{T}^{\mu\nu}_\mathbb{P}(t,q) &=& i 12 \sqrt{4\pi\alpha_{\rm em}}
\frac{m_{J/\psi}^2 \beta_c \beta_{c}}{f_{J/\psi}} \frac{1}{m_{J/\psi}^2-t} \left(
\frac{2\mu_0^2}{2\mu_0^2 + m_{J/\psi}^2 - t} \right) F_1(t)
\{ q\!\!\!/ \, g^{\mu\nu} - q^\mu \gamma^\nu \}\, ,\label{eq:pom}
\end{eqnarray}
where $t=(p_{N_i}-p_{N_f})^2$, $s=(q-p_{N_i})^2$,
$ \alpha_P (t) = \alpha_0 + \alpha'_P t$ with
$\alpha_0=1.25$ and $\alpha'_P=1/s_0=0.25$GeV$^{-1}$, $\mu_0^2=1.1$ GeV,
$\alpha_{\rm em} = e^2/4\pi=1/\sqrt{137}$,
$\beta_c=0.84$ GeV$^{-1}$,
 $f_{J/\psi}=11.2$, and $F_1(t) = (4M_N^2 - 2.8 t)/(4M_N^2 - t)(1-t/0.71(GeV^2))^2$.
For the $\pi$ exchange, the amplitude $\mathcal{M}^{\mu\nu}_\pi$ is:
\begin{eqnarray}
\mathcal{M}^{\mu\nu}_\pi(k,p_{N_f},q,p_{N_i})
&=& \frac{e}{f_\rho}\frac{g_{J/\Psi,\rho^0\pi^0}}{m_{J/\Psi}}\frac{f_{\pi NN}}{m_\pi}
\times \left(\frac{\Lambda^2}{\Lambda^2-t}\right)^4
\frac{1}{t-m^2_\pi} \epsilon^{\mu\nu\alpha\beta}k_\alpha q_\beta
[\gamma \cdot (p_{N_f}-p_{N_i})]\gamma^5\,,\nonumber\\
\label{eq:opep}
\end{eqnarray}
where the $f_\rho=5.33$, $g_{J/\Psi,\rho^0\pi^0}=0.032$,
and $\Lambda  = 2000$ MeV.

\begin{figure}[ht]
\centering
\epsfig{file=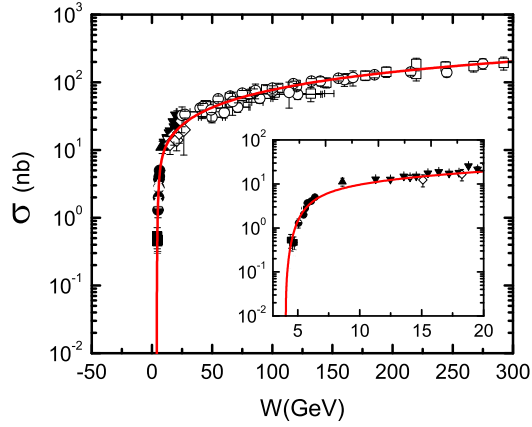, width=0.5\hsize}
\caption{The total cross section of $\gamma + p \to J/\psi + p$.
$W$ is the invariant mass of the $\gamma p$ system.
The red solid curves are calculated from the model
of Ref.\cite{wulee}.}
\label{fg:photonpm}
\end{figure}

All of the parameters specified above for evaluating
 Eqs.(\ref{eq:regge-g})-(\ref{eq:opep})
were determined in Ref.\cite{wulee}
by fitting the total cross section data of $\gamma + p \to J/\psi + p$
up to the invariant mass $W=300$ GeV, as shown in Fig.\ref{fg:photonpm}.

\subsubsection{$J/\psi N$ re-scattering amplitude}

The amplitude $\hat{T}^{J/\psi N}$ in Eq.(\ref{eq:diffcrsec})
( Fig.\ref{fg:diag}(c)) is :
\begin{eqnarray}
&&<\vec{k}\,m_{J/\psi},\;\vec{p}_p\;m_p,\;\vec{p}_n\;m_n|
\hat{T}^{J/\psi N}
|\vec{q}\;\lambda_{\gamma},\;-\vec{q}\;m_{d}>\nonumber\\
&=&\int d^3 \vec{k}^* \sum_{m_a,\;m_b\;=-1/2,\;1/2}\;\;\;
\sum_{m^*_{J/\psi}=-1,0,1}\nonumber\\
&&\times \left\{\;\;\; [ \frac{1}{W-E_N(\vec{p}_p)-E_{J/\psi}(\vec{k}^*)-E_N(-\vec{p}_p-\vec{k}^*)+i\epsilon}\right.\nonumber\\
&&\;\;\;\;\;\;\;\;\;\;\times<\vec{k}^*\;m^*_{J/\psi},\;\vec{p}_p\;m_p|
T_{J/\psi p,\gamma p}
|\vec{q}\;\lambda_\gamma,\;\vec{k}^*+\vec{p}_p-\vec{q}\;m_a>\nonumber\\
&&\;\;\;\;\;\;\;\;\;\;\times
<\vec{k}\,m_{J/\psi},\;\vec{p}_n\;m_n|
T_{J/\psi n,J/\Psi n}(W_{J/\Psi n})
|\vec{k}^*\;m^*_{J/\psi},\;\vec{k}+\vec{p}_n-\vec{k}^*\;m_b>\nonumber\\
&&\;\;\;\;\;\;\;\;\;\;\times
<\vec{k}^*+\vec{p}_p-\vec{q}\;m_a,\;\vec{k}+\vec{p}_n-\vec{k}^*\;m_b|
{\Phi}_{-\vec{q}\;\lambda_{d}}>]
+ [n \leftrightarrow p]\;\;\; \}
\label{eq:jpsiN}
\end{eqnarray}
where $W_{J/\Psi n}= E_{J/\Psi}(p_{J/\Psi}) + E_N(p_n)$,
$<\vec{p}\;m,\;\vec{p}^{\,\,'}\;,m'|\Phi_{\vec{p}_d;m_{d}}>$ has been
 defined by Eq.(\ref{eq:deut-wf}), and $ <\vec{k}\,m_{J/\psi},\;\vec{p}_n\;m_n|
T_{J/\psi n,J/\Psi n}(W_{J/\Psi n})
|\vec{k}^*\;m^*_{J/\psi},\;\vec{p}-\vec{k}^*\;m_b>$ can be calculated using
Eqs.(\ref{eq:cm-lab})-(\ref{eq:cceq-pw}).
The first term in the bracket denotes the $\gamma + p \to J/\psi + p$ and $J/\psi + n \to J/\psi + n$,
while the second term denotes $\gamma + n \to J/\psi + n$ and $J/\psi + p \to J/\psi + p$.

\subsubsection{$NN$ re-scattering amplitude}
The amplitude $\hat{T}^{NN}$ of Eq.(\ref{eq:diffcrsec})
(Fig.\ref{fg:diag}(b)) can be written as:
\begin{eqnarray}
&&<\vec{k}\,m_{J/\psi},\;\vec{p}_p\;m_p,\;\vec{p}_n\;m_n|
\hat{T}^{N N}
|\vec{q}\;\lambda_{\gamma},\;-\vec{q}\;m_{d}>\nonumber\\
&=&\int d^3\vec{p}^{\,\,*} \frac{1}{W-E_{J/\psi}(\vec{k})-E_N(\vec{p}^{\,\,*})
-E_N(-\vec{k}-\vec{p}^{\,\,*})+i\epsilon}\nonumber\\
&&\times\sum^{}_{m^*,\;m,\;m'\;=-1/2,\;1/2}
<\vec{k}\;m_{J/\psi},\;\vec{p}^{\,\,*}\;m^*|
T_{J/\Psi N,\gamma N}
|\vec{q}\;\lambda_\gamma,\;\vec{p}\;m>\nonumber\\
&&\times[
<\vec{p}_p\;m_p,\;\vec{p}_n\;m_n|
T_{np,np}(E_{np})
|\vec{p}^{\,\,*}\;m^*,\;\vec{p}^{\,\,'}\;m'>
<\vec{p}\;m,\;\vec{p}^{\,\,'}\;m'|
{\Phi}_{-\vec{q}\;m_{d}}>]\,,
\label{eq:NN}
\end{eqnarray}
where $\vec{p}^{\,\,'}=\vec{p}_p+\vec{p}_n-\vec{p}^{\,\,*}$
, $\vec{p}=\vec{k}+\vec{p}^{\,\,*}-\vec{q}$,
$E_{np}=E_N(p_p)+E_N(p_n)$, $<\vec{p};m,\;\vec{p}^{\,\,'};m'|\Phi_{\vec{p}_d;m_{d}}>$ has been
 defined by Eq.(\ref{eq:deut-wf}), and $<\vec{p}_p\;m_p,\;\vec{p}_n\;m_n|
T_{np,np}(E_{np})
|\vec{p}^{\,\,*}\;m^*,\;\vec{p}^{\,\,'}\;m'>$
can be calculated by using Eq.(\ref{eq:cm-lab-NN})
and the $NN\rightarrow NN$
partial wave amplitudes generated from the Bonn potential\cite{Machleidt}.

\subsection{Cross section of $\pi^+ + d \rightarrow J/\Psi + p + p$}
By replacing $T_{J/\Psi N,\gamma N}$ by $T_{J/\Psi N,\pi N}$ and
changing notations appropriately, the formula presented in the previous
subsection IV.B can be used to
calculate the differential cross section
$\frac{d\sigma}{d\Omega_p d|\vec{\kappa}_{J/\Psi}|}$
of $\pi^+ + d \rightarrow J/\Psi + p + p$.
\section{Results}

\begin{figure}[ht]
\centering
\epsfig{file=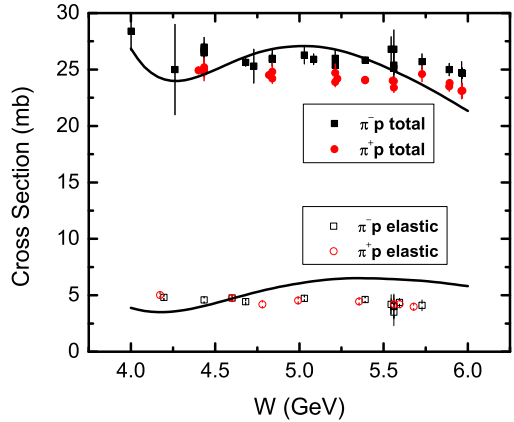, width=0.5\hsize}
\caption{The fits to the data of
 the total cross sections $\sigma^{tot}_{\pi^{\pm} p}$ of $\pi^{\pm} p$
reactions,
 and the total elastic $\pi^{\pm} p \rightarrow \pi^{\pm} p$
cross section $\sigma^{el}_{\pi^{\pm} p,\pi^{\pm}p}$. The data are from PDG\cite{pdg}}
\label{fg:pin-fit}
\end{figure}

\begin{figure}[ht]
\centering
\epsfig{file=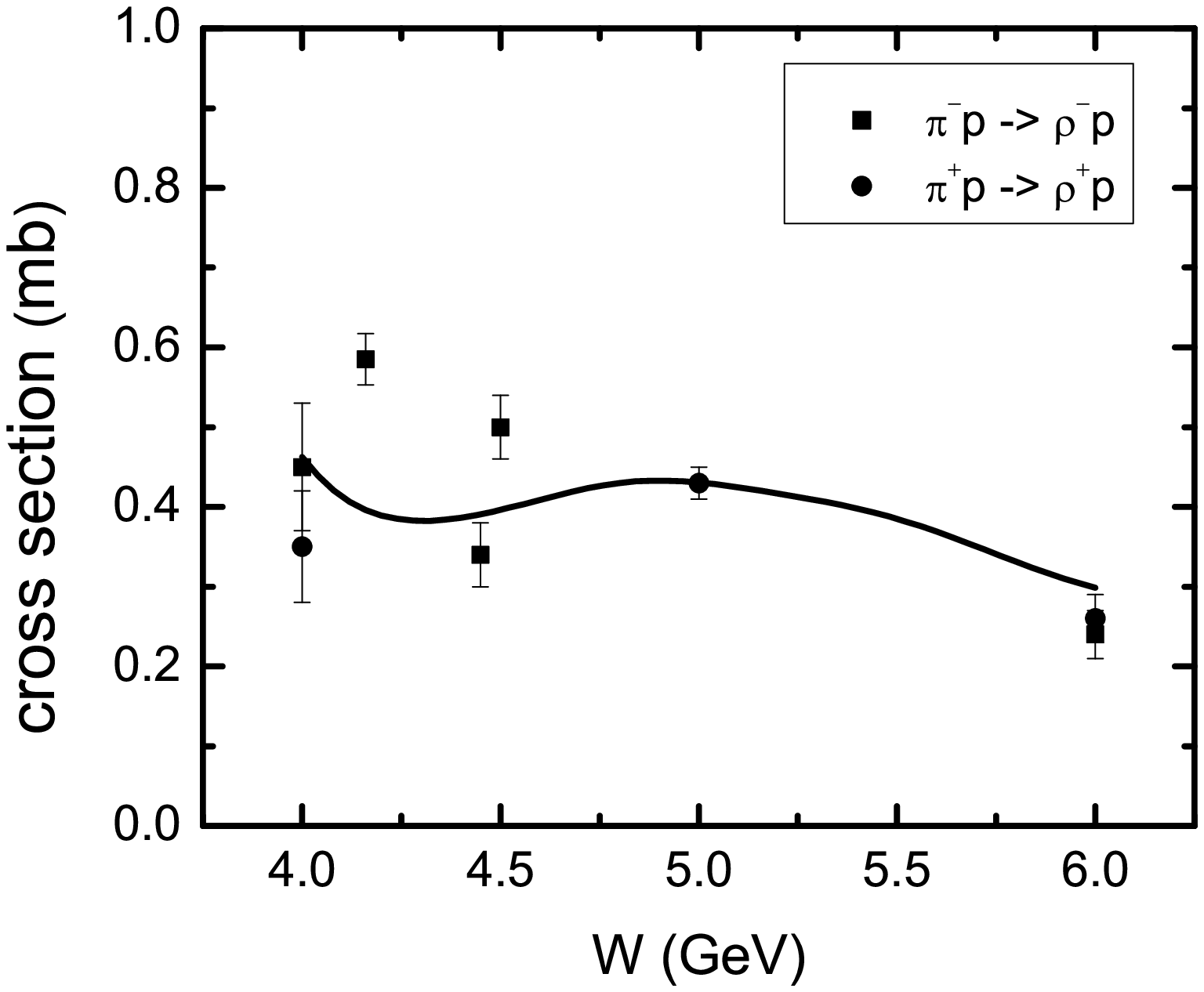, width=0.49\hsize}
\epsfig{file=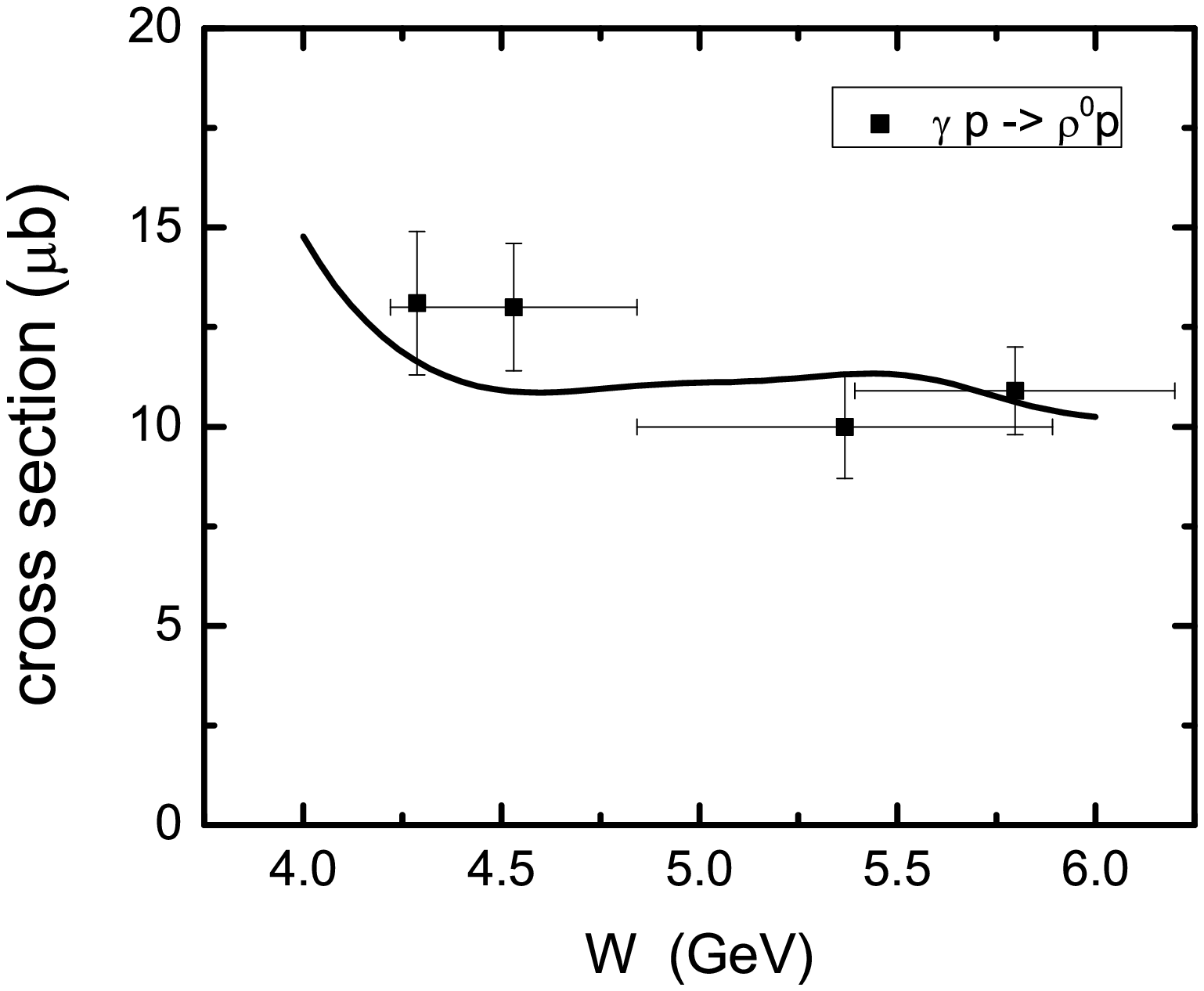, width=0.49\hsize}
\caption{The fits to the data of the total cross sections
 $\sigma_{\pi^{\pm} p \to \rho^{\pm}p}$ of
$\pi^{\pm}p \rightarrow \rho^{\pm} p$ (left)
 and $\sigma_{\gamma p \to \rho^0 p}$ for
$\gamma p \rightarrow \rho^0 p$ (right).
The data of $\sigma_{\pi^{\pm}p  \to \rho^{\pm} p}$ is from \cite{lbbook}, and the data of $\sigma_{\gamma p \to\rho^0 p}$ is from \cite{gprhopsjnp,gprhopprd,gprhopnpb}. }
\label{fg:rhon-fit}
\end{figure}

Our first task is to determine the parameters of
the potentials of the coupled-channel model
presented in section II. The parameters for the coupling
potential $V_{i,J/\Psi N}$
with $i=\pi N, \rho N, J/\Psi N$ have been specified there.
We determine the parameters of $V_{i,j}=v^0_{i,j}(E)f_{i,j}(r)$
with $i,j = \pi N,\rho N$, as already given
in Eq. (\ref{eq:v-fr}), by fitting the total cross section
 data  $\sigma^{tot}_{\pi N}$ of the $\pi N$ reaction,
$\sigma^{el}_{\pi N, \pi N}$ of $\pi N$ elastic scattering,
$\sigma_{\pi N, \rho N}$ of $\pi N \rightarrow \rho N$, and
$\sigma_{\gamma p,\rho^0 p}$ of $\gamma p \rightarrow \rho^0 p$.
We achieve this by choosing
\begin{eqnarray}
f_{i,j}(r)=\frac{1}{1+e^{(r-c)/t}
\label{eq:ws}}
\end{eqnarray}
where $c=0.8 $ fm, $t=0.4$ fm for all $i,j=\pi N,\rho N$.
The energy dependence of the total cross sections in the energy region
near the $J/\Psi$ production threshold ($4$ GeV $< W < $ 6 GeV)
can be fitted qualitatively by
setting
\begin{eqnarray}
v^0_{i,j}(E)=(-i)\;[A+B(E-E_0)^2]
\label{eq:v0ij}
\end{eqnarray}
Our fits to $\sigma^{tot}_{\pi^{\pm}p}$ are shown in
Fig.\ref{fg:pin-fit}. The small isospin dependence in the data is neglected in
our fits. In Fig.\ref{fg:rhon-fit}, we see that the fits
to the total cross section data
 of $\pi^{\pm}+ p \rightarrow \rho^{\pm} +p$ (left) and
$\gamma + p \rightarrow \rho^0 + p$ (right) are also reasonable.
The resulting parameters $A$, $B$, and $E_0$ are listed in Table \ref{tab:v}.

Note that the forms Eqs.(\ref{eq:ws})-(\ref{eq:v0ij}) are purely
phenomenological, and our fits in Figs.\ref{fg:pin-fit}-\ref{fg:rhon-fit}
are only qualitatively. But they are sufficient for estimating the order
of magnitudes of the $J/\Psi$ production cross sections. For a more quantitative
calculation, we clearly need to improve this phenomenological aspect of
our model.

With the fits shown in Figs.\ref{fg:pin-fit}-\ref{fg:rhon-fit}, all of the
parameters for our calculations are completely fixed. In the next few
subsections, we present our predictions for future experimental determinations
of the $J/\Psi$-$N$ interaction.

\begin{table}[ht]
\begin{center}
\caption{The parameters  for $v^0_{i,j}(E)$ of Eq.(\ref{eq:v0ij}).
$E$ is the
total energy in the enter of mass system.}
\label{tab:v}
\begin{tabular}{cccccc}\hline
      i  &       j     & A (GeV)      & B (GeV$^{-1}$)&$E_0$ (GeV) \\
$\pi N$  &  $\pi N$   & 0.43        &   1.18 & 4.0 \\
$\pi N$  &  $\rho N$  & 0.04        &   0.09 & 4.0\\
$\rho N$ &  $\rho N$  & 0.45        &   0.45 & 4.0\\
\hline
\end{tabular}
\end{center}
\end{table}

\subsection{$\pi^- + p \rightarrow J/\Psi + n$}
\begin{figure}[ht]
\centering
\epsfig{file=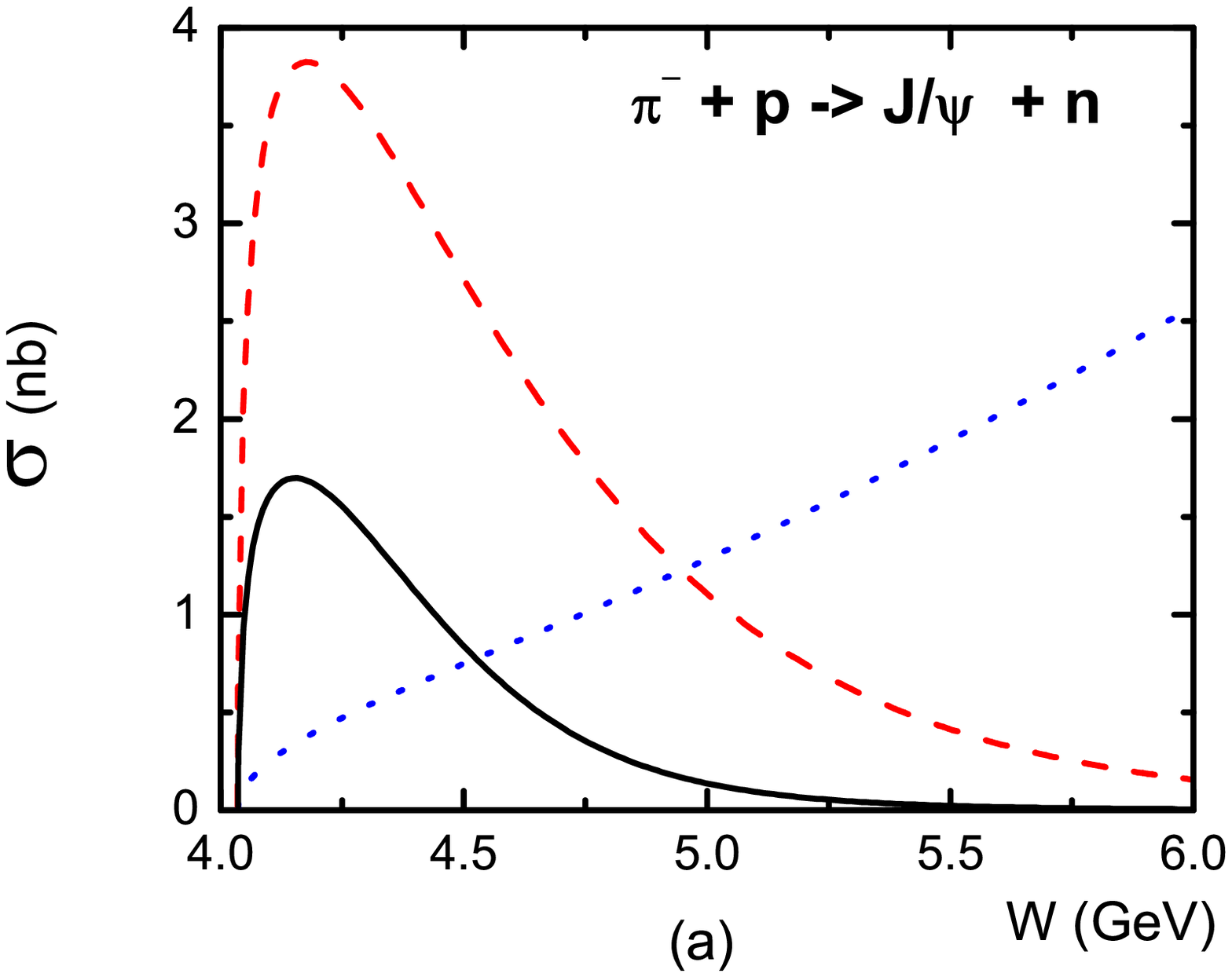, width=0.49\hsize}
\epsfig{file=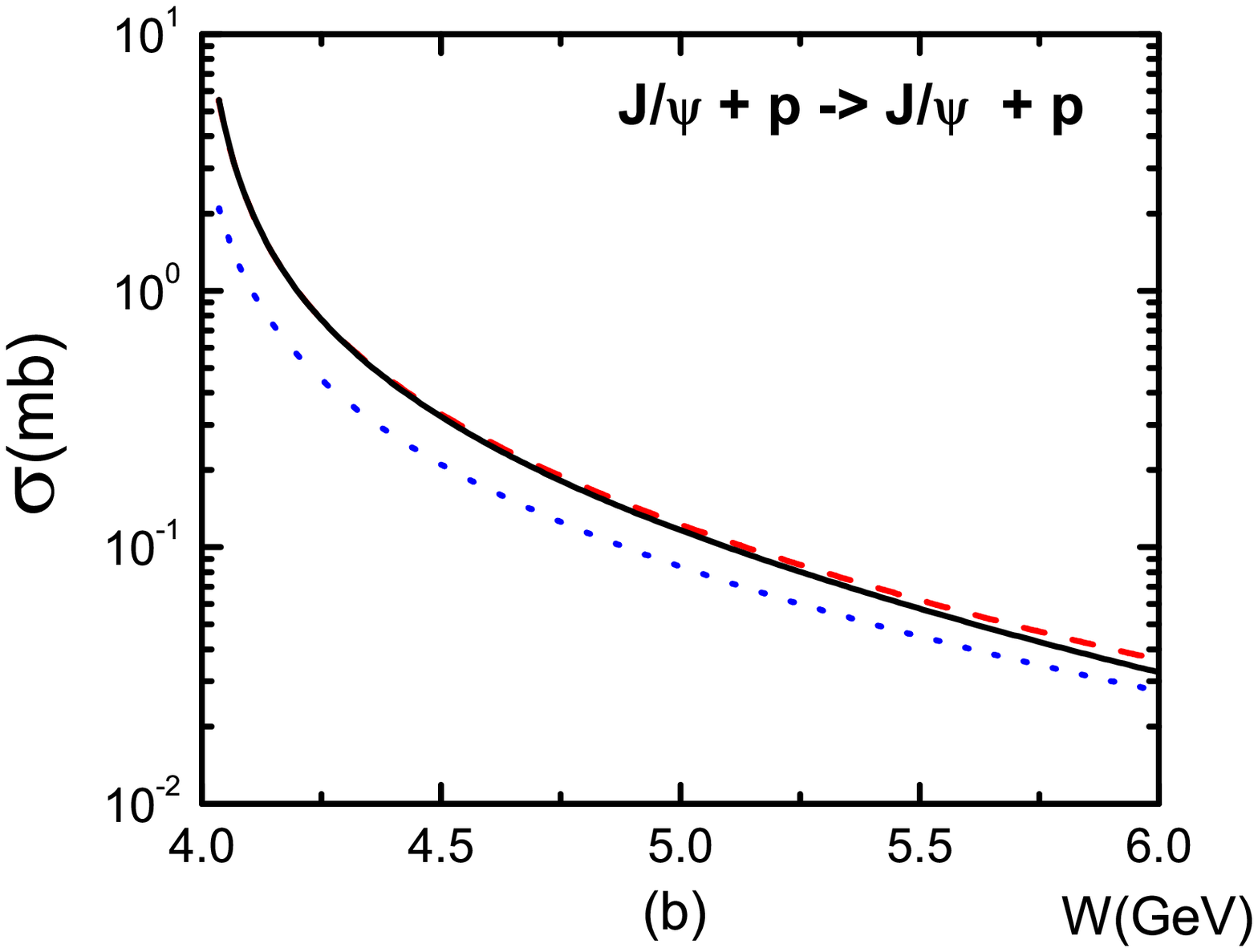, width=0.49\hsize}
\caption{ The cross sections
of $\pi^- + p \rightarrow J/\Psi + n$ (left)
 and $J/\Psi + p\rightarrow J/\Psi + p$ (right).
Black Solid: full calculation,
red dashed:  Coupled-channel, but setting
 $V_{i,J/\Psi}=0$ for $i,j=\pi N, \rho N$,
blue dotted : Born calculation $T_{i,j}= V_{i,j}$. W is the total energy in
the center of mass frame. }
\label{fg:pin-cc}
\end{figure}

The simplest experiment to determine the $J/\Psi$-$N$ interaction within our
coupled-channel model is to measure the
cross sections of $\pi^- + p \rightarrow J/\Psi + n$  reaction.
We first observe that the coupled-channel effects can
change drastically the magnitudes and the energy-dependence
of the predicted cross sections. This is illustrated
in the left side of  Fig.\ref{fg:pin-cc}.
In the absence of coupled-channel effects, the results from the Born
approximation (setting
 $T_{J/\Psi N, \pi N}= V_{J/\Psi N, \pi N}$) are
the dotted blue
curve.  When the interactions associated with the $J/\Psi N$ channel,
$V_{i,J/\Psi}$ with $i=\pi N, \rho N, J/\Psi N$, are included
in solving the coupled-channel equation, we obtain
the dashed red curve which is suppressed at high $W$ in contrast to
the raising behavior of the dotted blue
curve from the Born approximation.
When the complex potentials $V_{i,j}$ with $i,j= \pi N, \rho N$ are
also included in
our full calculations, we obtain the solid black curve.
We see that these complex potentials can reduce drastically
the magnitudes of the cross sections. This is due to the fact that most of
the incident pions are absorbed before the $J/\Psi$ production takes place.
Within our model, this absorption effect is due to the very large imaginary
part of $V_{\pi N, \pi N}$.

The coupled-channel effects on
 $J/\Psi+p \rightarrow J/\Psi +p$ are shown in the right side of
Fig.\ref{fg:pin-cc}. We note that the $J/\Psi$-$N$ interaction, as defined in
Eq.(\ref{eq:vjpnjpn}), is real and therefore does not have strong absorption effects
on  $J/\Psi$-$N$ scattering. By comparing the differences between the blue
dotted curve of the Born approximation ($T_{J/\Psi N, J/\Psi N}=V_{J/\Psi N, J/\Psi N}$)
 and the red dashed curve,
we see that the coupled-channel
effects due to $V_{i,J/\Psi}$ with $i=\pi N, \rho N, J/\Psi N$ is to
increase the cross section at all $W$. The coupled-channel effects due to
 $V_{i,j}$ with $i,j= \pi N, \rho N$ are very small, as can be seen by comparing
the red dashed curve and the black solid curve in the right side of
Fig.\ref{fg:pin-cc}. This is due to the fact that
the  $J/\Psi$-$\pi$-$\rho$ coupling constant $g_{J/\Psi\rho\pi}=0.032$, calculated from
the partial decay width of $J/\Psi\rightarrow\pi\rho$, is very small
in the calculation of Eqs.(\ref{eq:pinjn})-(\ref{eq:rhonjn}) for the one-meson-exchange
matrix elements of
$v_{\rho N,J/\Psi}$  and $v_{\pi N,J/\Psi}$.


\begin{figure}[ht]
\centering
\epsfig{file=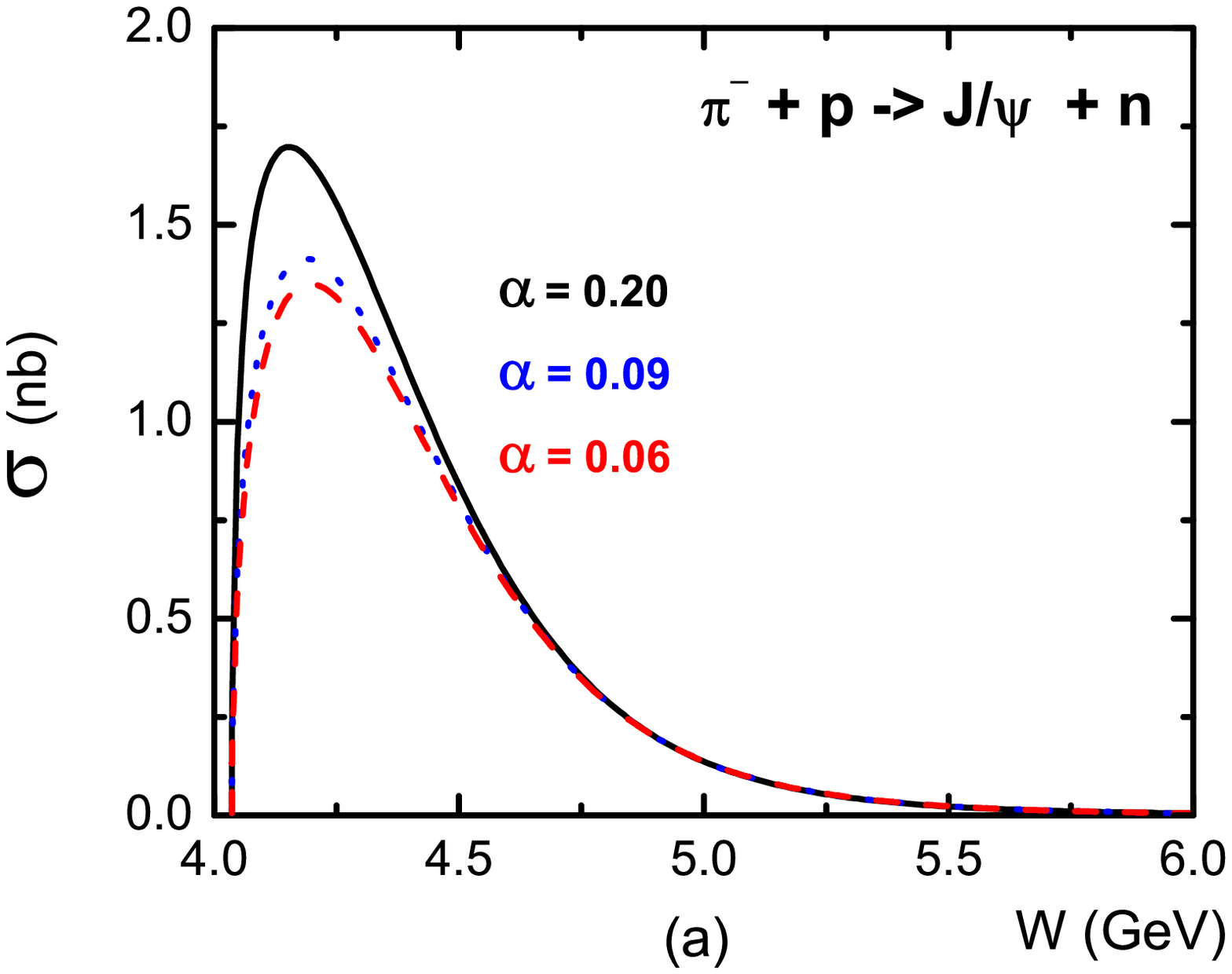, width=0.49\hsize}
\epsfig{file=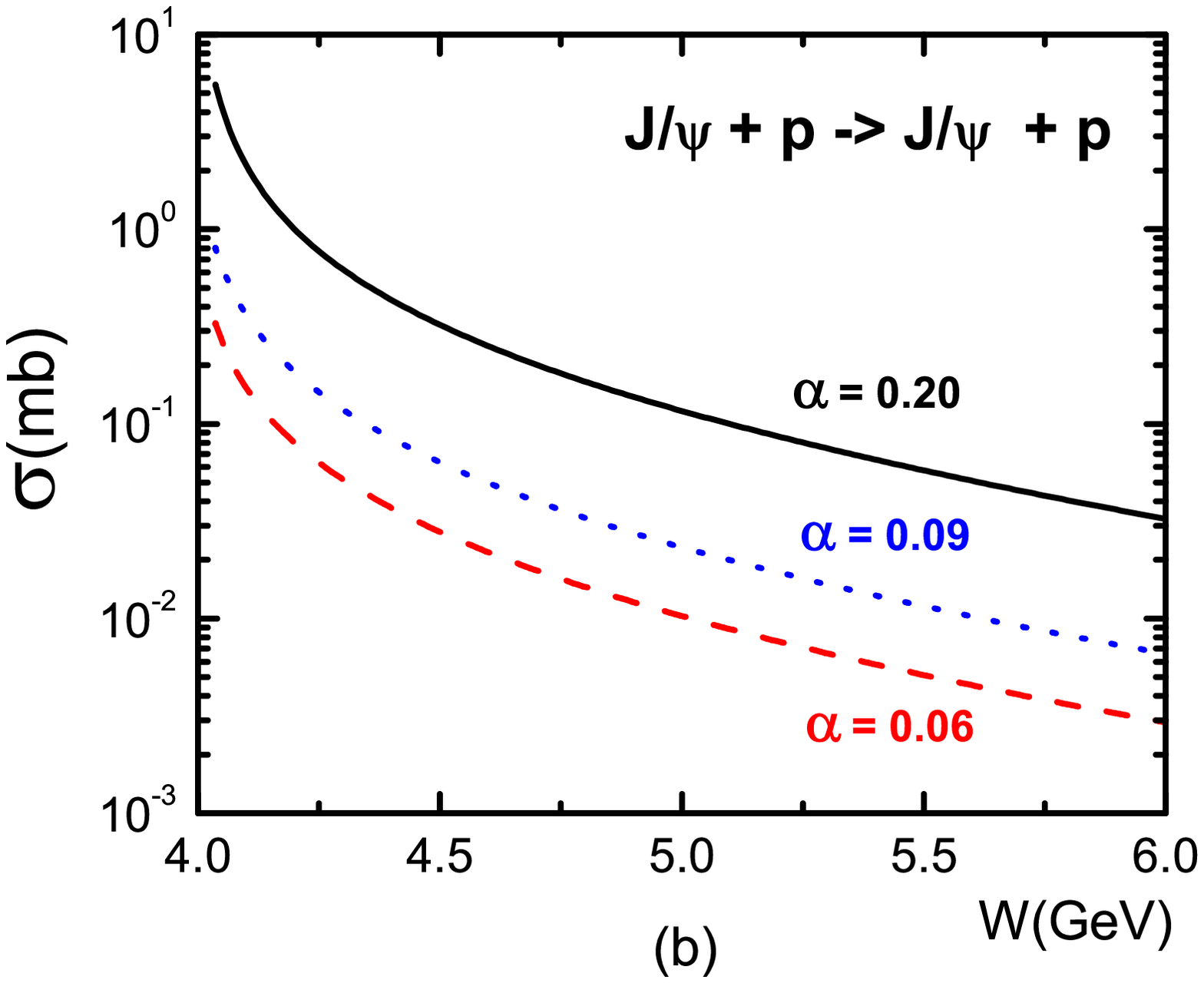, width=0.49\hsize}
\caption{The total cross sections of
 $\pi^- + p \to J/\Psi + n$ (left) and
$J/\psi + p \to J/\Psi + p$ (right).
W is the total energy in the center of mass frame.
The black solid, blue dotted, and red dashed curves are calculated
by using
the $J/\Psi$-$N$ potential Eq.(\ref{eq:vjpnjpn}) with
 $\mu=0.6$ GeV and $\alpha =0.20, 0.09,0.06$, respectively. }
\label{fg:pin-jpsin}
\end{figure}

We next examine the dependence of the $\pi^-+ p \rightarrow J/\Psi + n$
cross sections on the strength $\alpha$ of the $J/\Psi$-$N$ potential
$v_{J/\Psi N, J/\Psi N}$.
The
results from using $\alpha =0.2, 0.09, 0.06$ are compared in
the left side of Fig.\ref{fg:pin-jpsin}.
 The predicted total cross
sections at peak positions are about 1.5 nb. Such small cross sections are due to the
strong absorption of the incident pion and that the $\pi N \rightarrow J/\Psi N$ transition
potential $V_{\pi N,J/\Psi N}$ is weak, as discussed above.
The differences between three
results are significant only in the region near $W \sim 4.25$ GeV.
On the other hand, the predicted $J/\Psi N \rightarrow J/\Psi N$
cross sections are more sensitive to $\alpha$, as seen in the
right side of Fig.\ref{fg:pin-jpsin}. This suggests that
$J/\Psi$-$N$ interaction can be more easily determined in the
reactions of $J/\Psi$ production on the  deuteron target in the kinematic region
where the cross sections are
sensitive to the $J/\Psi N \rightarrow J/\Psi N$
re-scattering mechanism (Fig.\ref{fg:diag}(c)).
This is what we will examine  in
the next two subsections.

\subsection{$\gamma + d \rightarrow J/\Psi +n + p$}

We only consider the energy region close to
the $J/\Psi$ production
threshold. The results presented below  and in the next subsection
are calculated
at the energy 100 MeV above the $J/\Psi$ production.

\begin{figure}[ht]
\centering
\epsfig{file=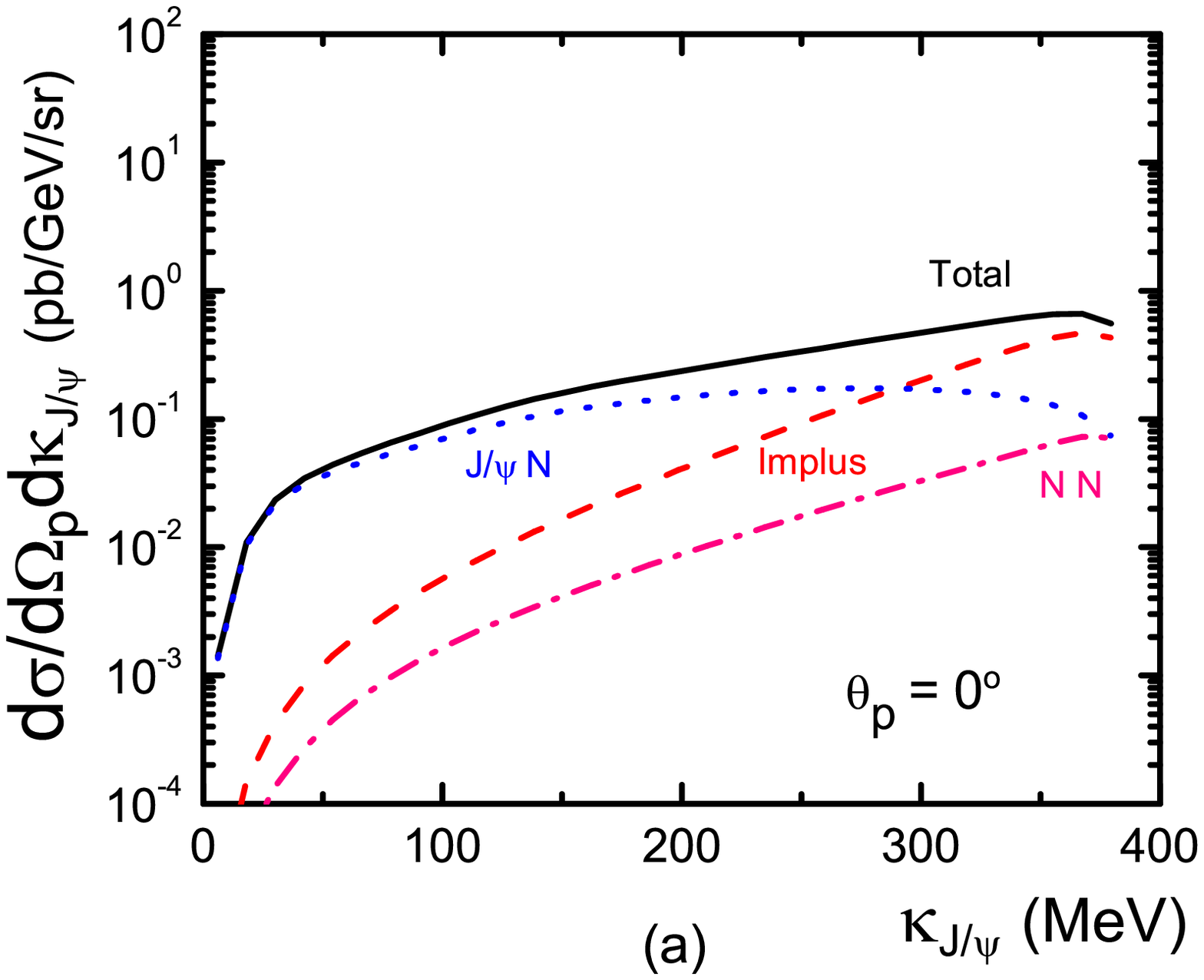, width=0.49\hsize}
\epsfig{file=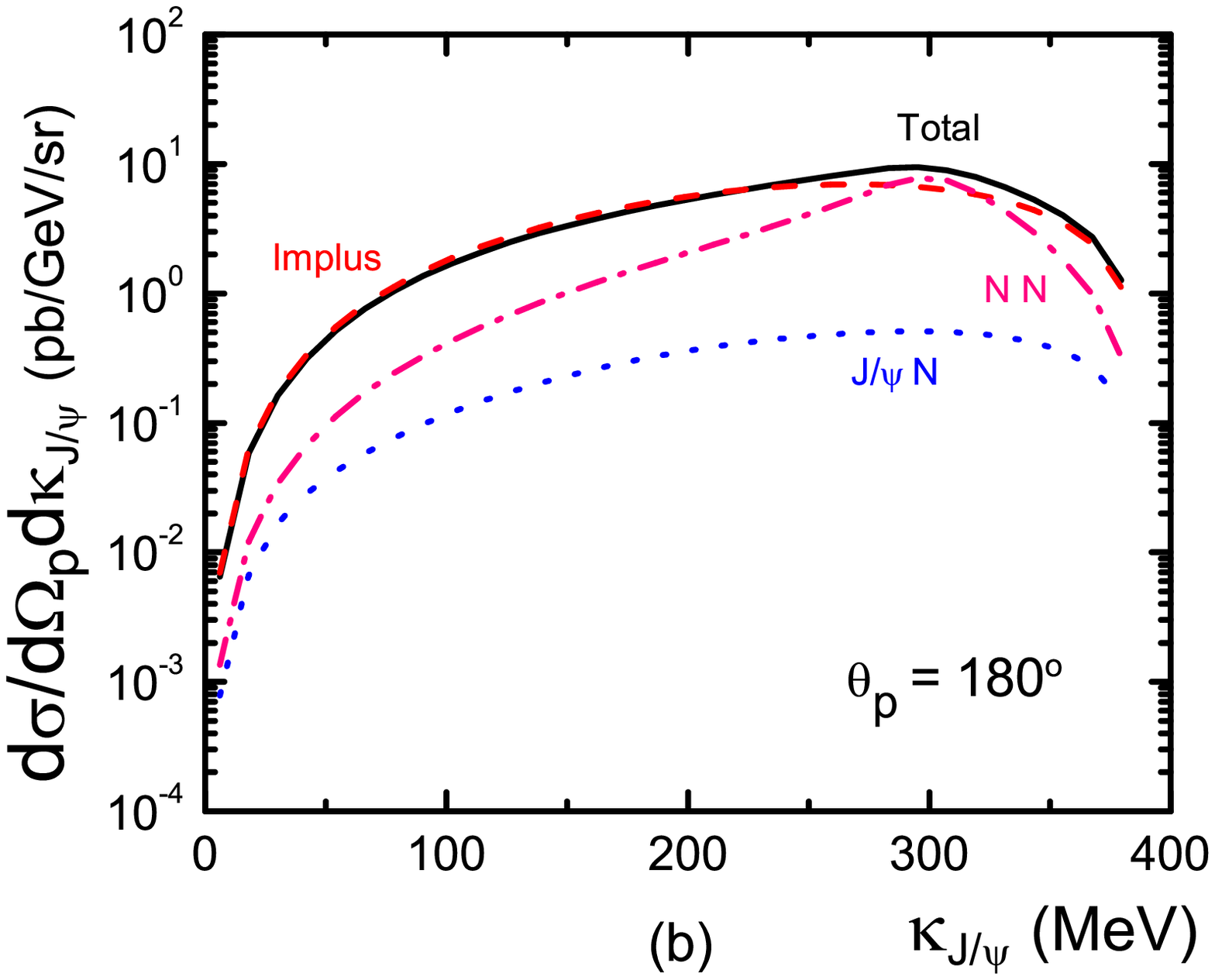, width=0.49\hsize}
\caption{The differential cross section $d\sigma/(d\Omega_p d\kappa_{J/\Psi})$
of $\gamma + d \to J/\Psi + p + n$ vs the momentum $\kappa_{J/\Psi}$
 of $J/\Psi$ in
 the center of mass frame of the $J/\Psi + n$ sub-system.
The results are from coupled-channel calculations
 using the $J/\Psi$-$ N$ potential,
Eq.(\ref{eq:vjpnjpn}), with
$\mu=0.6$ GeV and $\alpha =0.20$. $\theta_p$ in
 (a) and (b) are  the angle between the incoming photon and the outgoing proton.
The red dashed, blue dotted and pink dashed-dotted lines are
the contributions from the amplitudes of
the Impulse term, $J/\Psi N$ re-scattering, and $N N$ re-scattering, respectively.
The black solid lines are the coherent sum of these three amplitudes.}
\label{fg:phodpm}
\end{figure}

\begin{figure}[ht]
\centering
\epsfig{file=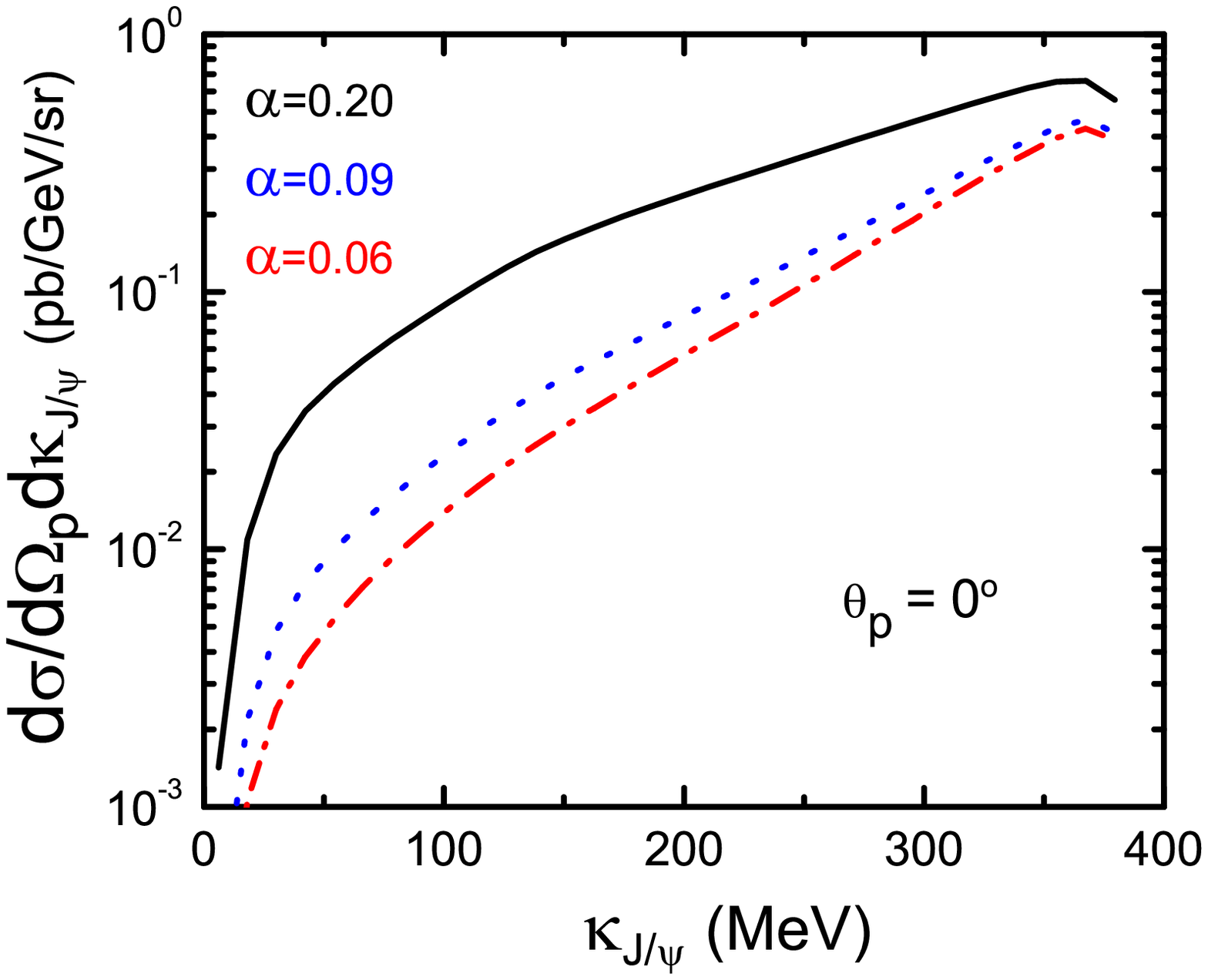, width=0.49\hsize}
\epsfig{file=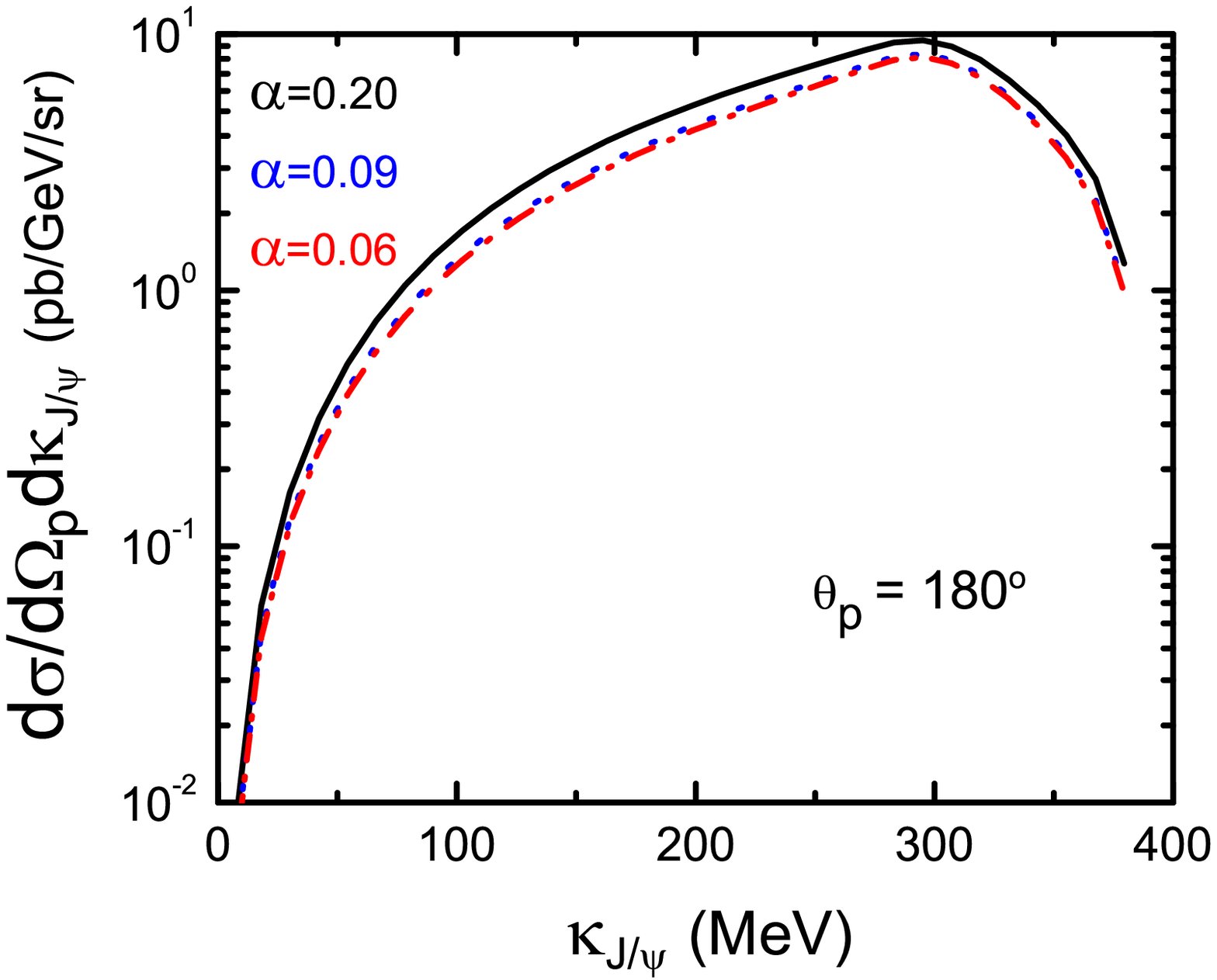, width=0.49\hsize}
\caption{
The differential cross section $d\sigma/(d\Omega_p d\kappa_{J/\Psi})$
of $\gamma + d \to J/\Psi + p + n$ vs the momentum $\kappa_{J/\Psi}$
 of $J/\Psi$ in
 the center of mass frame of the $J/\Psi + n$ sub-system.
$\theta_p$
 is  the angle between the incoming photon and the outgoing proton.
 The black solid, blue dotted, and red dashed-dotted lines are from
the coupled-channel calculations  using
the $J/\Psi N \to J/\Psi N$ potential, Eq.(\ref{eq:vjpnjpn}), with
 $\mu=0.6$ GeV and $\alpha =0.20, 0.09,0.06$, respectively.}
\label{fg:phodjn}
\end{figure}

As discussed in section IV.D, our main task is to use  Eq.(\ref{eq:diffcrsec})
 to
examine the dependence of  the differential cross section
$\frac{d\sigma}{d\Omega_pd|\vec{\kappa}_{J/\Psi}|}$ of this process on the $J/\Psi$-$N$ potential
$V_{J/\Psi N, J\Psi N}$.
Obviously,
we  need to
 identify the region of the outgoing proton angle
$\theta_p$ with respect to the incident photon where the
$J/\Psi$-$N$ re-scattering mechanism ((c) of Fig.\ref{fg:diag}) dominates.
We find that this is in the region close to $\theta_p=0$, as
can be seen in the left side of Fig.\ref{fg:phodpm}. We see
that in the low $J/\Psi$ momentum,
$\kappa_{J/\Psi} <$ about 200 MeV, the contribution from
the $J/\Psi$ re-scattering term (blue dotted curve) is much larger than that from the
other two
mechanisms. On the other hand, the impulse term dominates at $\theta_p =180^0$ as seen in
the right side of Fig.\ref{fg:phodpm}.
Accordingly, three $J/\Psi$-$N$ potentials with the strengths
$\alpha=0.2$, $0.09$, and $0.06$ obtained by the effective field theory approach and LQCD
can be easily tested at $\theta_p=0$, but less
 possible at $\theta_p=180^0$.
This is shown in Fig.\ref{fg:phodjn}.

The results presented above clearly indicate that
 it is highly desirable to have data in the region close to $\theta_p=0$
in future experiments in order to be able to determine the $J/\Psi$-$N$ interaction.

\subsection{$\pi^+ + d \rightarrow J/\Psi + p + p$}
\begin{figure}[ht]
\centering
\epsfig{file=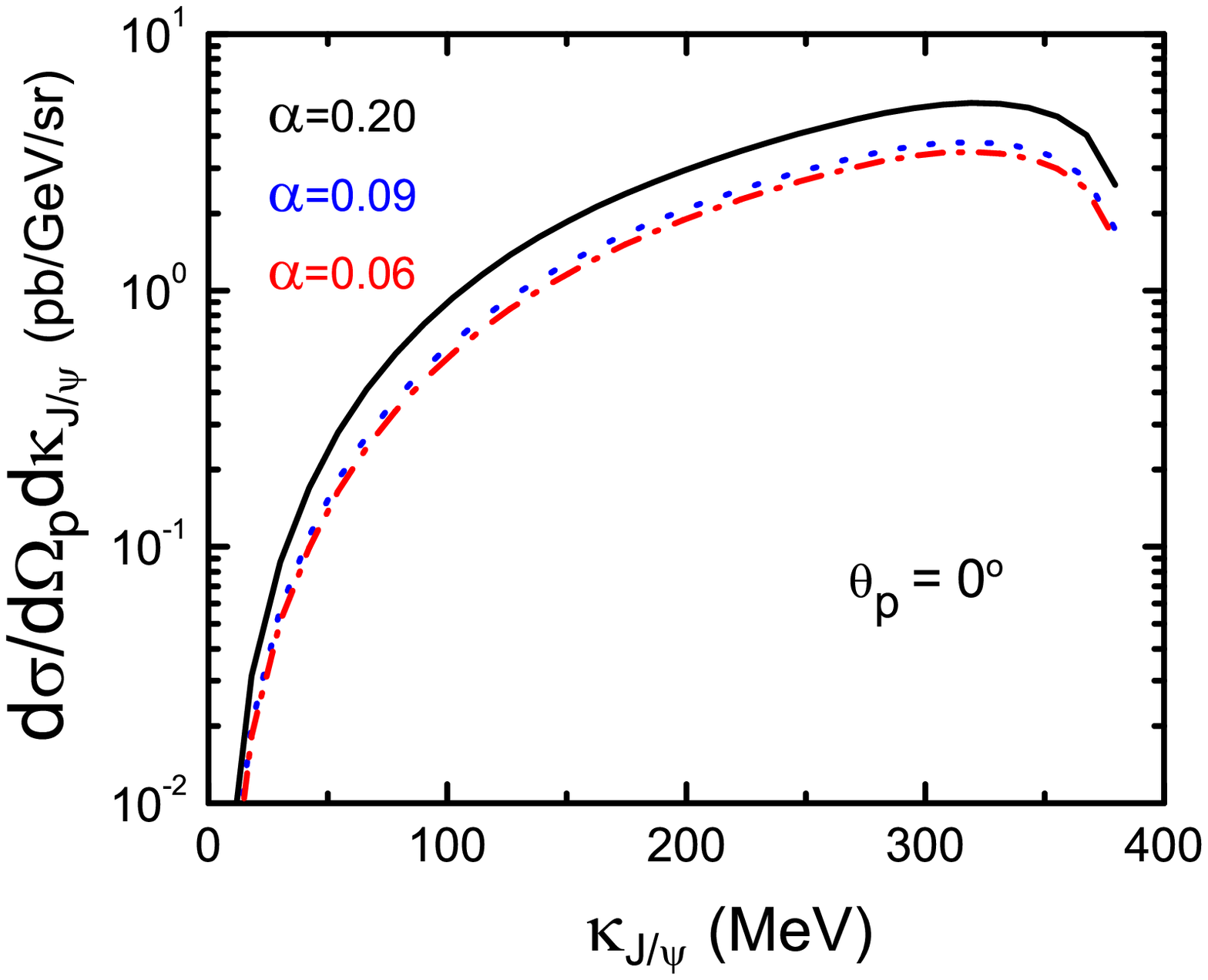, width=0.49\hsize}
\epsfig{file=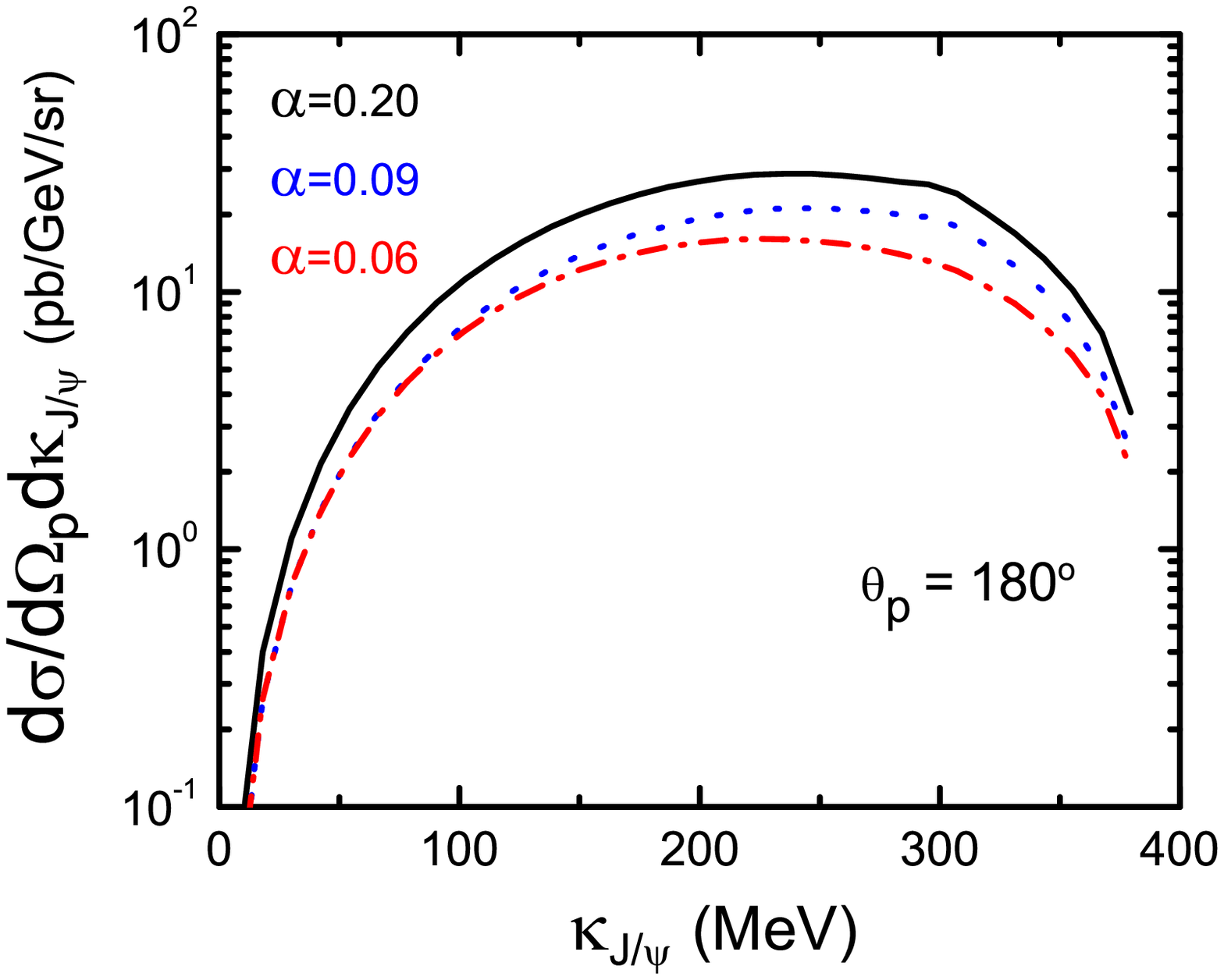, width=0.49\hsize}
\caption{
The differential cross section $d\sigma/(d\Omega_p d\kappa_{J/\Psi})$
of $\pi^+ + d \to J/\Psi + p + p$ vs the momentum $\kappa_{J/\Psi}$
 of $J/\Psi$ in
 the center of mass frame of the $J/\Psi + p$ sub-system.
The black solid, blue dotted, and red dashed-dotted lines are from
the coupled-channel calculations
 using the $J/\Psi$-$ N$ potential,
Eq.(\ref{eq:vjpnjpn}), with
$\mu=0.6$ GeV and $\alpha =0.20, 0.09, 0.06$. $\theta_p$ is
the angle between the incoming pion and the outgoing proton.
 }
\label{fg:pinpid}
\end{figure}

For the calculated $\frac{d\sigma}{d\Omega_p d|\vec{\kappa}_{J/\Psi}|}$ of
the $\pi^+ d \to J/\Psi + p + p$ reaction, we find that
 the impulse term dominates at all angles. Furthermore,
the contributions from the $J/\Psi$-$N$ re-scattering (Fig.\ref{fg:diag}(c))
are weaker than those of
the $NN$ re-scattering (Fig.\ref{fg:diag}(b)).
Nevertheless, the data from experiments on
this reaction  can be useful
to determine the $J/\Psi$-$N$ interaction.
This is illustrated in
 Fig.\ref{fg:pinpid} where  the results from
$\alpha =0.20$, $0.09$, and $0.06$
for $\theta_p=0$ and $180^0$ are compared.

\subsection{$\gamma p \rightarrow J/\Psi p$ in the coupled-channel model}
\begin{figure}[ht]
\centering
\epsfig{file=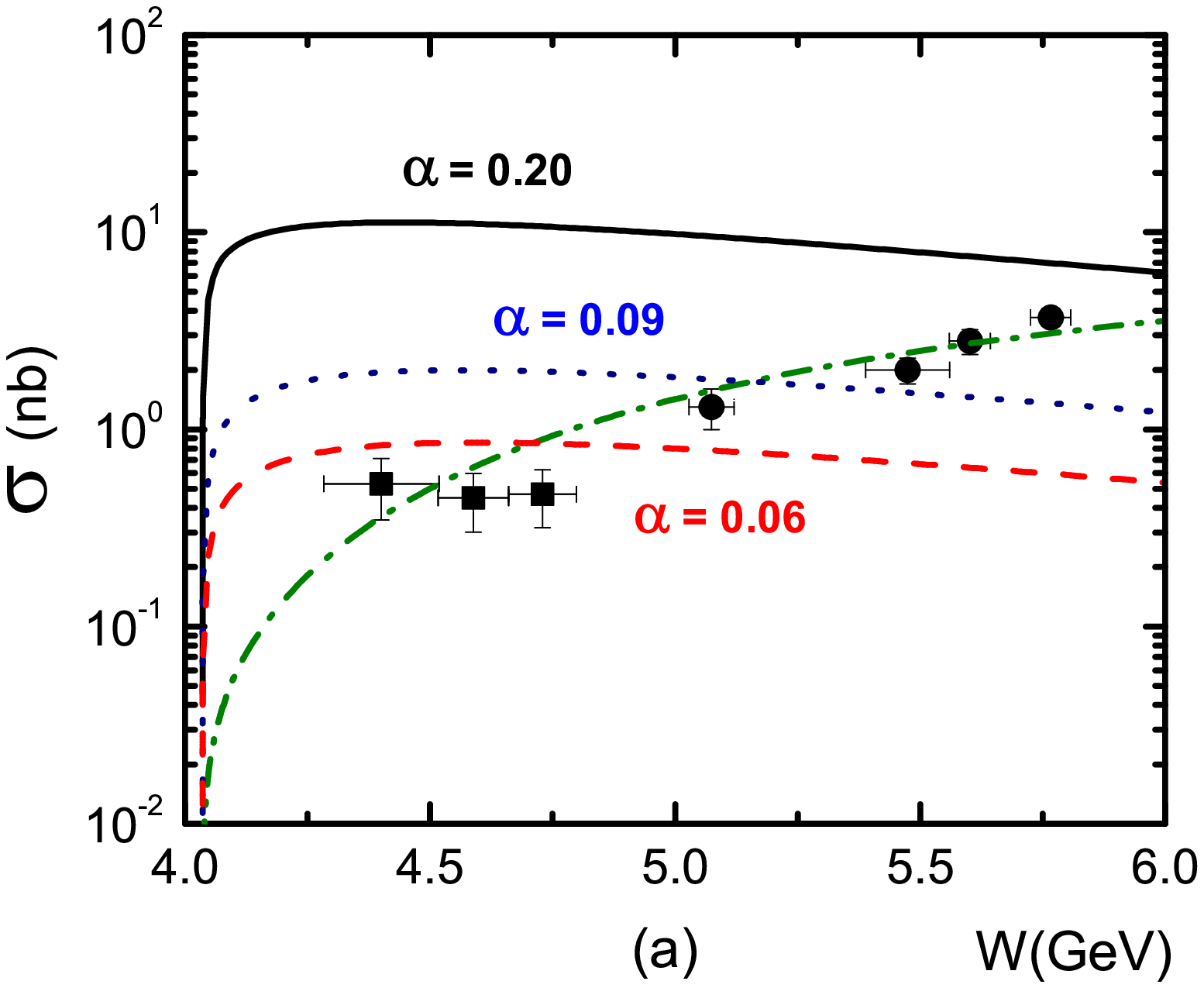, width=0.49\hsize}
\epsfig{file=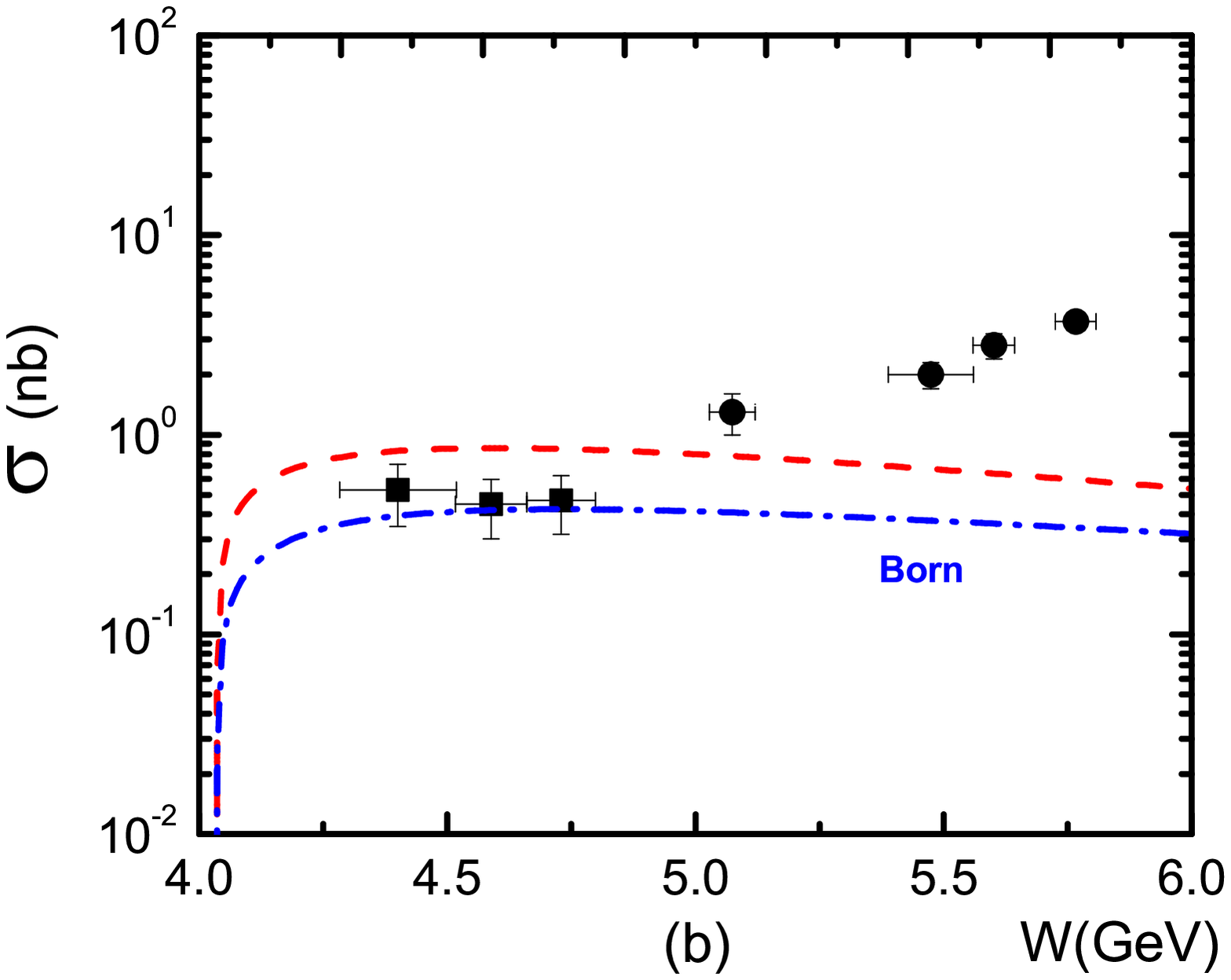, width=0.49\hsize}
\caption{The cross section of $\gamma + N \to J/\Psi + N$ reaction.
$W$ is the total energy in the center of mass frame.
In the left side, the black solid, blue dotted, red dashed lines are calculated by using  couple channel model with
$\mu=0.6$ GeV and $\alpha =0.20, 0.09,0.06$ and the green dashed-dotted line is from Pomeron-exchange.
In the right side, the red dashed and blue dashed-dotted lines are from the full calculation and
the Born approximation  calculation $T_{i,j}= V_{i,j}$.
The parameters of $v_{J/\Psi N,J/\Psi N}$ of the
coupled-channel model are $\mu=0.6$ GeV and $\alpha =0.06$. The data is from \cite{Gittelman:1975ix,Camerini:1975cy,Anderson}.}
\label{fg:phpphd}
\end{figure}

With the vector meson dominance hypothesis, we can predict the cross section of
$\gamma + p \rightarrow J/\Psi+ p$
within the constructed coupled-channel model.
This is done by using the amplitude given in
Eq.(\ref{eq:gn-jn}).
For three coupled-channel amplitudes calculated with
$\alpha = 0.2$, $0.09$, and $0.06$  for $V_{J/\Psi N, J/\Psi N}$, our results
 are compared with that
of the Pomeron-exchange model (green dot-dashed curve)
in the left of Fig.\ref{fg:phpphd}.

We note that the coupled-channel model results are expected to be valid
only at energies near $J/\Psi$ production threshold where the data are rather
uncertain.
On the other hand, the use of Pomeron-exchange model at very low energies
is also questionable.
Our results for $\alpha=0.06$ are close to the three data points near 4.5 GeV. At $W >$ about 5 GeV, we clearly need to improve the model.
 The data from the forthcoming experiment\cite{jpsi-jlab}
 at JLab
will be useful to clarify the situation.

In the right side of Fig.\ref{fg:phpphd}, we show that the coupled-channel effects
can  increase the cross section significantly. Our results here as well as that shown in
Fig.\ref{fg:pin-cc} suggest that
any attempt to determine the $J/\Psi$-$N$ interaction must include the coupled-channel
effects, as required by the unitarity condition.

\section{Summary}

We have developed a coupled-channel model
with $\pi N$, $\rho N$ and $J/\Psi N$ channels
to predict the $J/\Psi$ production on the nucleon.
The $J/\Psi$-$N$ interaction potential $V_{J/\Psi N,J/\Psi N}$ is taken from
the calculations using the effective field theory method and Lattice QCD.
The $J/\Psi N \rightarrow \pi N, \rho N$ transition potentials
$V_{J/\Psi N, \pi N}$ and $V_{J/\Psi N, \rho N}$ are calculated
from one-$\rho$-exchange and one-$\pi$-exchange mechanisms with
the parameters determined by the decay width of
$J/\Psi \rightarrow \pi \rho$ and the previously determined $\pi NN$ and
$\rho NN$ coupling constants. The interactions
$V_{i,j}$ with $i,j=\pi N, \rho N$ are treated as phenomenological
complex potentials with their parameters constrained by the fits to
the data of total cross sections of the $\pi N$ reactions, $\pi N\rightarrow \pi N$,
$\pi N \rightarrow \rho N$, and
$\gamma  p \rightarrow \rho^0 p$.

The calculated meson-baryon amplitudes
are then used to predict the cross sections
of $\gamma + d \rightarrow J/\Psi + N + N$ and
$\pi + d \rightarrow J/\Psi + N + N$ reactions.
The calculations on the deuteron target
involve the contributions from the
impulse term and the final $NN$ and $J/\Psi N$ scattering.
 We have identified the kinematic region where
the  $J/\Psi$-$N$ potentials can be distinguished
 in $\pi^-+p \rightarrow J/\Psi + n$, $\gamma + d \rightarrow J/\Psi + n + p$
and $\pi^++ d \rightarrow J/\Psi +p+p$ reactions.
Predictions of the dependence of the cross sections of
these reactions on
the $J/\Psi$-$N$ potentials are presented. Our results shown in Figs.\ref{fg:pin-jpsin},
\ref{fg:phodjn}, and \ref{fg:pinpid} can be used to facilitate the experimental
determinations of $J/\Psi$-$N$ interactions.

Within the vector meson dominance model, we have also applied the constructed
coupled-channel model to predict the $\gamma + p\rightarrow J/\Psi + p$
cross sections. Our results near the $J/\Psi$ production threshold, as shown in
Fig.\ref{fg:phpphd},
are very different from what can be calculated from  the
conventional Pomeron-exchange model which is mainly constrained by
the $\gamma p \rightarrow J/\Psi + p$ cross section at high energies.
It will be interesting to distinguish these two models in the forthcoming
experiment at Jefferson Laboratory\cite{jpsi-jlab}.

\begin{acknowledgments}
We thank R. Machleidt for providing us with the code for generating the $NN$ amplitudes
from the Bonn potential.
This work is supported by the U.S. Department of Energy, Office of Nuclear Physics Division,
under Contract No. DE-AC02-06CH11357.
This research used resources of the National Energy Research Scientific Computing Center,
which is supported by the Office of Science of the U.S. Department of Energy
under Contract No. DE-AC02-05CH11231, and resources provided on ``Fusion,''
a 320-node computing cluster operated by the Laboratory Computing Resource Center
at Argonne National Laboratory.
\end{acknowledgments}

\end{document}